\documentclass[12pt]{article}
\pdfoutput=1
\usepackage{epsf, cite, amssymb}
\usepackage{epsfig}
\usepackage{amsmath}
\usepackage[all]{xypic}

\makeatletter
\renewcommand\section{\@startsection {section}{1}{\z@}%
                                   {-3.5ex \@plus -1ex \@minus -.2ex}
                                   {2.3ex \@plus.2ex}%
                                   {\normalfont\large\bfseries}}
\renewcommand\subsection{\@startsection{subsection}{2}{\z@}%
                                     {-3.25ex\@plus -1ex \@minus -.2ex}%
                                     {1.5ex \@plus .2ex}%
                                     {\normalfont\bfseries}}
\makeatother

\let\non\nonumber

\let\w=\wedge

\newcommand{\del}{\partial}

\def\one{^{(1)}}

\def\inv{^{-1}}

\newcommand{\bea}{\begin{eqnarray}}
\newcommand{\eea}{\end{eqnarray}}
\newcommand{\be}{\begin{equation}}
\newcommand{\ee}{\end{equation}}
\newcommand{\bma}{\begin{pmatrix}}
\newcommand{\ema}{\end{pmatrix}}
\newcommand{\cc}{{\rm c.c.}}

\newcommand{\Z}{{\mathbb Z}}
\newcommand{\R}{{\mathbb R}}

\newcommand{\PP}{{\mathbb P}}
\newcommand{\CC}{{\mathbb C}}

\newcommand{\cM}{{\cal M}}
\newcommand{\B}{{\cal B}}

\newcommand{\wt}{\widetilde}
\newcommand{\wh}{\widehat}

\newcommand{\SO}{\operatorname{SO}}

\newcommand{\SL}{\operatorname{SL}}

\newcommand{\SU}{\operatorname{SU}}

\newcommand{\U}{\operatorname{U}}
\renewcommand{\O}{\operatorname{O}}

\newcommand{\e}{\epsilon}
\newcommand{\ve}{\varepsilon}

\newcommand{\C}[1]{$(\ref{#1})$}

\def\IZ{\relax\ifmmode\mathchoice
{\hbox{\cmss Z\kern-.4em Z}}{\hbox{\cmss Z\kern-.4em Z}}
{\lower.9pt\hbox{\cmsss Z\kern-.4em Z}} {\lower1.2pt\hbox{\cmsss
Z\kern-.4em Z}}\else{\cmss Z\kern-.4em Z}\fi}
\def\IR{\relax{\rm I\kern-.18em R}}

\def\one{{\hbox{ 1\kern-.8mm l}}}

\def\bz{{\bar z}}
\def\bw{{\bar w}}

\def\Tr{{\rm Tr\,}}

\newlength{\bredde}
\def\slash#1{\settowidth{\bredde}{$#1$}\ifmmode\,\raisebox{.15ex}{/}
\hspace*{-\bredde} #1\else$\,\raisebox{.15ex}{/}\hspace*{-\bredde}
#1$\fi}

\newsavebox{\zzzbar}
\sbox{\zzzbar}
  {\setlength{\unitlength}{0.9em}
  \begin{picture}(0.6,0.7)
  \thinlines
  \put(0,0){\line(1,0){0.6}}
  \put(0,0.75){\line(1,0){0.575}}
  \multiput(0,0)(0.0125,0.025){30}{\rule{0.3pt}{0.3pt}}
  \multiput(0.2,0)(0.0125,0.025){30}{\rule{0.3pt}{0.3pt}}
  \put(0,0.75){\line(0,-1){0.15}}
  \put(0.015,0.75){\line(0,-1){0.1}}
  \put(0.03,0.75){\line(0,-1){0.075}}
  \put(0.045,0.75){\line(0,-1){0.05}}
  \put(0.05,0.75){\line(0,-1){0.025}}
  \put(0.6,0){\line(0,1){0.15}}
  \put(0.585,0){\line(0,1){0.1}}
  \put(0.57,0){\line(0,1){0.075}}
  \put(0.555,0){\line(0,1){0.05}}
  \put(0.55,0){\line(0,1){0.025}}
  \end{picture}}

\def\Im{{\rm Im ~}}

\newcommand{\ena}{\end{eqnarray}}
\newcommand{\beqa}{\begin{eqnarray}}
\newcommand{\eeqa}{\end{eqnarray}}

\newcommand{\half}{\frac{1}{2}}

\def\H{{\cal H}}

\newcommand{\zbar}{{\bar z}}

\newfont{\goth}{ygoth.tfm scaled 1200}                   

\def\e{\epsilon}

\def\O{\Omega}

\renewcommand{\SO}{\operatorname{SO}}
\renewcommand{\O}{{\mathcal{O}}}
\renewcommand{\U}{\operatorname{U}}

 \numberwithin{equation}{section}

\def\1{{(1)}}
\def\2{{(2)}}
\def\3{{(3)}}

\def\1{{\bf 1}}

\def\B{{\mathcal B}}

\def\M{{\mathcal M}}

\begin{document}
\begin{titlepage}

\begin{center}

{30 April 2010}
\hfill         \phantom{xxx} \hfill EFI-09-26

{revised: 28 October 2010}
\hfill DAMTP-2010-33

\hfill UCSB-Math-2010-09

\vskip 2 cm {\Large \bf Geometries, Non-Geometries,  and Fluxes}\non\\
\vskip 1.25 cm { Jock McOrist$^{a}$\footnote{j.mcorist@damtp.cam.ac.uk},
David R. Morrison$^{b}$\footnote{drm@math.ucsb.edu}, and
Savdeep Sethi$^{c, d}$\footnote{sethi@uchicago.edu}}\non\\
{\vskip 0.5cm $^{a}$ Department of Applied Mathematics and
Theoretical Physics, Centre for Mathematical Sciences, Wilberforce
Road, Cambridge, CB3 OWA, UK\non\\ \vskip 0.2 cm $^{b}$ Departments
of Mathematics and Physics, University of California, Santa Barbara,
CA 93106, USA\non\\ \vskip 0.2 cm $^{c}$ Enrico Fermi Institute,
University of Chicago, Chicago, IL 60637, USA\non\\ \vskip 0.2cm
$^{d}$ Institute for Theoretical Physics, University of Amsterdam,
Valckenierstraat 65, 1018 XE Amsterdam, The Netherlands \non\\}

\end{center}
\vskip 2 cm

\begin{abstract}
\baselineskip=18pt

Using F-theory/heterotic duality, we describe a framework for
analyzing non-geometric $T^2$-fibered heterotic compactifications to
six- and four-dimensions. Our results suggest that among
$T^2$-fibered heterotic string vacua, the non-geometric
compactifications are just as typical as the geometric ones.
 We also construct four-dimensional solutions
 which have novel type IIB and M-theory dual descriptions. These
 duals are non-geometric
  with  three- and four-form fluxes not of
$(2,1)$ or $(2,2)$ Hodge type, respectively, and yet preserve at
least $N=1$ supersymmetry.

\end{abstract}

\end{titlepage}

\pagestyle{plain}
\baselineskip=19pt
\newpage
\tableofcontents
\newpage


\section{Introduction}
\subsection{The basics of non-geometries} \label{thebasics}
The space of four-dimensional string compactifications is
potentially vast. The degeneracy of these vacua comes about by the
many choices of compactification metric and associated fluxes. When
the size of the compactification space is large compared with the
string scale, we can use supergravity to study the resulting
low-energy four-dimensional physics. However, we expect generic
stabilized vacua to involve string scale physics for which
supergravity is inadequate.

One way in which a compactification space can become quantum is if
the patching conditions involve symmetries present in string theory
but not supergravity. The simplest example of this type is F-theory
where the backgrounds involve seven-brane sources of type IIB string
theory~\cite{Vafa:1996xn}. Without knowing that S-duality is a good
symmetry of type IIB string theory, those backgrounds would make no
sense as solutions of type IIB supergravity. A second example of
quantum patching conditions are compactifications that involve
T-duality, aspects of which we will explore here. This second case is an example of quantum geometry which arises in classical string theory, much like mirror symmetry.

\begin{figure}
\includegraphics[scale=.8]{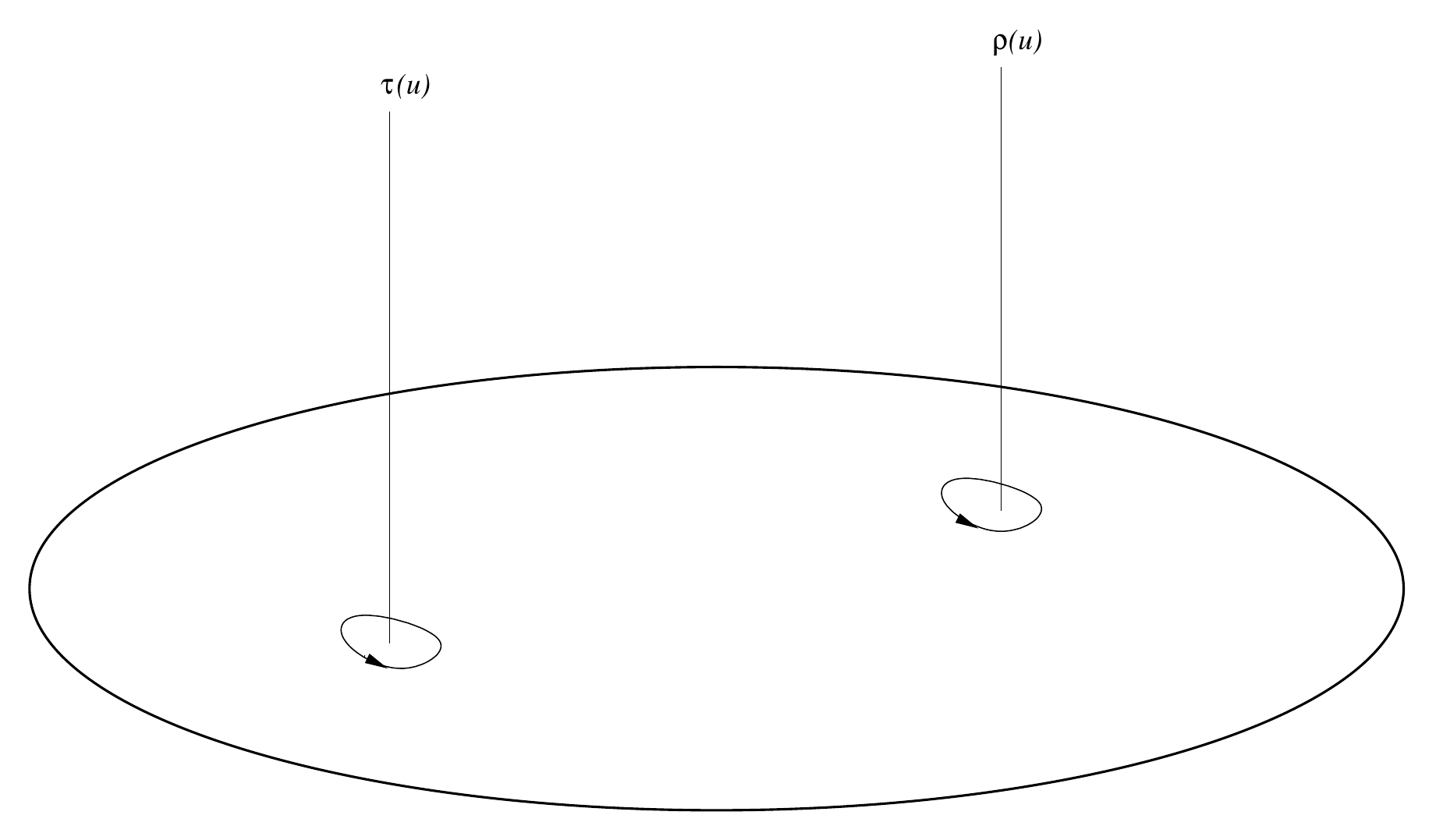}\label{fibration}
\caption{\it A schematic of the desired fibration data where $u$
denotes coordinates on the base $B$. The loci of $\tau$ and $\rho$
degenerations can be viewed as supporting $5$-branes.}
\end{figure}

Closed string theory on $T^2$ has two basic moduli: the complex
structure parameter $\tau$ of $T^2$ and the K\"ahler modulus $\rho$
which determines the volume $V$ of $T^2$ and the $\B$-field, \be \label{definerho} \rho
= \rho_1 + i\rho_2= \B+iV. \ee To build an elliptic compactification,
one usually fibers $\tau$ over a base space allowing $\tau$ to
undergo monodromies valued in $SL(2,\Z)$. These are large
diffeomorphisms of the torus. In string theory, however, $\tau$ and
$\rho$ share the same symmetry group, appearing on equal footing and
we should be able to describe quantum compactifications where both
$\tau$ and $\rho$ vary over a base space as depicted in
figure~\ref{fibration}. Since the action of $SL(2,\Z)$ on $\rho$ includes $V \rightarrow
1/V$, these compactifications are typically inherently quantum. This is the class of compactifications we wish to explore.

 In the purely geometric case where a large volume limit is possible, we
can describe a torus fibration over a base $B$,  depicted in
figure~\ref{fibration}, using a local semi-flat approximation for
the metric \be \label{fiberedmetric} ds^2 = g_{ij} du^i du^j +
{\frac{\rho_2}{\tau_2}} \, | dw_1 + \tau(u) dw_2|^2. \ee The base
metric is $g_{ij}$ and the torus has coordinates $(w_1, w_2)$. The
complex structure $\tau(u)$ varies over $B$ while $\rho$ is
constant. This metric possesses $U(1)\times U(1)$ isometries acting
on the torus fibers. For compact spaces, the semi-flat
metric~\C{fiberedmetric}\ is typically used as an approximation to a smooth
Ricci flat metric with no isometries, with the approximation becoming exact as $V\rightarrow 0$. It is the
smooth metric which is used to define the world-sheet sigma model, which flows to a conformal field theory defining the perturbative string background.
 However, with both $\tau$ and
$\rho$ varying, the existence of a smooth metric is no longer
possible. Consequently, the condition analogous to the existence of a smooth metric should be the existence of a conformal field theory
specified by $\tau, \rho$ and $B$.

If we reduce $10$-dimensional  string theory on $T^2$ to
$8$-dimensions then we can view the resulting theory as possessing
$2$ families of $(p,q)$ $5$-branes in analogy with the $(p,q)$
$7$-branes of type IIB string theory. One family is associated with
$\tau$ degenerations while the other is associated with $\rho$
degenerations. From this perspective, compactifications on $B$, like
the one in figure~\ref{fibration}, include $5$-branes at the
degeneration loci of $\tau$ and $\rho$. The standard NS5-brane
corresponds to a purely perturbative $\rho$ monodromy. If the moduli of the compactification can be tuned to make all the
$\rho$ monodromies perturbative then the model is likely to admit an
asymmetric orbifold description. This is analogous to the
orientifold limit of F-theory proposed by Sen~\cite{Sen:1997gv}.

The most desirable approach for studying stringy compactifications
involving ingredients like T-duality is a  world-sheet analysis
where $\alpha'$ effects can be determined directly. In type II
string theory this kind of analysis can be further complicated by
the presence of Ramond-Ramond (RR) fields, branes and orientifolds.
These ingredients, needed for N=1 compactifications with stabilized
moduli~\cite{Dasgupta:1999ss}, are difficult to analyze beyond the
large volume supergravity limit, though it may be possible to
understand their role in the Berkovits
formalism~\cite{Berkovits:2002zk}; see, for
example~\cite{Linch:2008rw}.

In contrast, the heterotic string is a more desirable framework to
use for two key reasons. Firstly, solutions are specified purely by
the Neveu-Schwarz (NS) field content, which consists of the metric,
torsion flux and bundle data. This avoids many of the complications
of RR fluxes and, in principle, it is possible to construct
world-sheet descriptions of heterotic vacua within the RNS
formalism. Secondly, no orientifolds are needed. The Bianchi
identity for the ${\cal H}_3$-flux of the heterotic string, \be
\label{bianchi} d {\cal H}_3 = { \frac{\alpha'} 4} \left( \Tr (R
\wedge R) - \Tr (F \wedge F) \right), \ee automatically includes a
higher derivative curvature term that makes compact solutions
possible. This removes the typically difficult task of consistently
patching orientifold actions together globally with T-dualities.
This makes it much easier to construct non-geometric heterotic
solutions than type II or M-theory solutions. We will see how this
simplifies the description of non-geometric vacua in the heterotic
string versus type IIB orientifolds in
sections~\ref{hettorsion}-\ref{iibm}.

One of the aims of this paper is to make use of heterotic--F-theory duality to provide a purely
geometric description of a large class of non-geometric heterotic
compactifications. The duality is typically stated as follows:
F-theory compactified on a $K3$-fibered Calabi-Yau $(n+1)$-fold,
which is also elliptically-fibered with section is equivalent to the heterotic string compactified on an
elliptically fibered Calabi-Yau $n$-fold.
Usually, one takes a particular limit in the moduli space of elliptic
$K3$ surfaces to ensure that the heterotic solution is at large volume
and well-described by supergravity.

From the point of view of F-theory, there is nothing special about
this point in the moduli space, and one can ask what happens more
generally. In this paper, refining some work of Clingher and Doran
\cite{math.AG/0602146}, we extend the heterotic--F-theory duality beyond
the traditional limit, to all points in the moduli space
where the heterotic gauge group remains unbroken.

As we will see, the generic heterotic solution with a dual F-theory
description may not have a large volume limit but may instead
involve patching by the T-duality group of the heterotic string on
$T^2$. This provides a very nice way of determining fibration data
for non-geometric compactifications. In fact, the F-theory fibration
captures not only $\tau$ and $\rho$ but also the Wilson line data
for the heterotic gauge bundle on $T^2$.\footnote{In geometric
models where $V$ can be made arbitrarily large, this Wilson line
data describes a flat $(E_8\times E_8)\rtimes \mathbb Z_2$
connection on $T^2$.} So this approach should lead to the
construction and description of quantum bundles. Exact conformal 
field theory descriptions of local heterotic models with abelian bundles 
have been found in~\cite{Carlevaro:2008qf, Carlevaro:2009jx}. 
It would be very interesting to see if that approach
can be extended to accommodate non-geometric bundles. 

It is important to stress that for compactifications  with N=1
supersymmetry, the F-theory/heterotic duality is not generally a
quantum equivalence of string vacua. Rather, it is a means by which
we can obtain classical data to describe a heterotic
compactification. In the geometric case  (without ${\cal
H}_3$-flux), this data is an elliptic Calabi-Yau space over $B$ and
a holomorphic bundle which provides the defining data for a
heterotic sigma model. In the non-geometric case, this data is
replaced by a fibration of $\tau$ and $\rho$ over $B$ and a quantum
bundle. However, aside from special BPS couplings, most space-time
quantities such as K\"ahler potentials are going to be different in
each theory.

\subsection{Beyond $T^2$ and other approaches}

It is natural to expect this geometrization of quantum heterotic
compactifications to extend beyond $T^2$ fibrations. Indeed, if most
Calabi-Yau spaces can be described as $T^3$-fibered spaces, as
conjectured by Strominger--Yau--Zaslow~\cite{syz}, then we should expect ``generic''
heterotic compactifications to involve patching by the quantum
symmetry group of the heterotic string on $T^3$ whose moduli space
involves several distinct components~\cite{triples}. The quantum
patching conditions or monodromy data of the $T^3$-fibration should
then be captured by M-theory compactified on a (potentially
singular) $K3$-fibered manifold with $G_2$ holonomy. This is
important to understand if we are to enumerate string vacua.
Unfortunately, little is known about the construction of compact
$G_2$ spaces let alone spaces admitting $K3$-fibrations so we will
restrict our attention to heterotic compactifications with
$T^2$-fibrations.

The final interesting case is a $T^4$-fibered heterotic
compactification. In this case, we expect the quantum heterotic
compactification to admit a dual description in terms of type IIA on
a $K3$-fibered space which also involves quantum patching conditions
(namely, mirror transforms of the $K3$ fiber). In this case, both
sides of the duality are generically quantum.

Some of the first attempts to construct quantum compactifications
using U-duality appear in~\cite{Kumar:1996zx}. The type II examples
considered were compactifications to three dimensions or lower
mainly because the solutions involved the full U-duality group
rather than subgroups like the T-duality group. This work also
pre-dates the discovery of flux vacua and the associated more
general metric ans\"atze like the non-K\"ahler solutions
of~\cite{Dasgupta:1999ss}. These more general metrics will be
important in the examples we construct in section~\ref{hettorsion}.

More recently, a detailed discussion of non-geometric type II
solutions in six dimensions appeared in~\cite{Hellerman:2002ax}. The
type II construction involves fibering $T^2\times T^2$ which gives a
double elliptic fibration over a base. The torus factors capture the
$\tau$ and $\rho$ monodromies. This doubled torus formalism has been
further discussed in~\cite{Hull:2004in}\ where backgrounds using
T-duality in the patching conditions have been termed ``T-folds."
The doubled torus approach has been extended to the heterotic string
very recently in~\cite{ReidEdwards:2008rd}.

This doubled torus approach should be contrasted with the geometry
of a $K3$-fibration that we use here. In principle, one should be
able to understand global properties like tadpole cancellation from
the doubled torus formalism but it looks less intuitive for the
heterotic string. This is partly because the definition of both
sides of the Bianchi identity~\C{bianchi}\ are unclear, and partly
because the bundle plays an important role in solving the tadpole
condition~\C{bianchi}; that bundle data is naturally encoded in the
$K3$ fibration. For N=1 compactifications, the tadpole conditions
are really quite critical. For type II non-geometric backgrounds,
there are similar issues which remain to be
understood~\cite{Hellerman:2002ax}. 

The doubled torus approach might, however, be useful for
constructing world-sheet descriptions; see, for
example~\cite{Avramis:2009xi}. For example, it might be possible to extend the 
beta function computation of the doubled torus sigma model, developed 
in~\cite{Berman:2007xn}, to derive a complete version of the tadpole condition discussed in 
section~\ref{tadpoles}.  That is a quite critical issue.

Our approach suggests a very
different heterotic world-sheet description obtained naturally by
studying an M5-brane wrapped on the $K3$-fiber of the dual geometry.
Such an M5-brane sigma model can capture both torsional and
torsion-free geometries along the lines discussed
in~\cite{Sethi:2007bw}. We plan to explore this interesting wrapped
brane configuration elsewhere. The last approach that leads
naturally to non-geometric backgrounds is T-dualizing flux vacua.
This approach was explored, for example, in~\cite{Shelton:2006fd}.
For a review of past work on non-geometric backgrounds,
see~\cite{Wecht:2007wu}.

\subsection{Some open issues and an outline}

Some of the basic outstanding questions for non-geometric compactifications
can be summarized as follows:
\begin{itemize}
\item What fibration data is needed to describe such compactifications?
\item How do we construct and analyze world-sheet models which involve quantum patching conditions?
\item What new phenomenology or low-energy physics is possible in this wider class of compactifications?
\end{itemize}
We will set up a framework to answer the first two points. It would be very interesting to extend this framework
beyond $T^2$ heterotic fibrations to $T^3$ fibrations. The third
question is also extremely interesting. At least in type II models, it
appears that new low-energy couplings do emerge from non-geometric
compactifications as described in~\cite{Ihl:2007ah}. It seems
reasonable to suspect that new phenomenology might emerge in
heterotic compactifications as well.

Most of the heterotic backgrounds we will describe are not
left-right symmetric on the world-sheet. To describe a type II
compactification, we would like to know if an analogue of the
standard embedding exists with varying $\rho$. It seems reasonable
that such a generalization exists and will provide type II solutions
in a way quite different from the U-manifold geometrization
discussed in~\cite{Kumar:1996zx}.

Lastly, there should be nice methods of taking these solutions and
generating non-geometric heterotic solutions {\it without} F-theory
dual descriptions. For example, in the geometric setting,
quotienting an ellipic Calabi-Yau with section by a free action can
result in a torus-fibered Calabi-Yau without a section. The
resulting space is still perfectly fine for the heterotic string but
no longer fits into the heterotic/F-theory duality framework. We
expect analogous constructions for these non-geometric models.

The outline for the paper is as follows: we first
reconsider heterotic--F-theory duality in section~\ref{f-het},
focusing on the case of unbroken heterotic gauge group. Our
analysis leads to a new construction of non-geometric heterotic
compactifications in section~\ref{NGHM}.  The
solutions we describe will be primarily phrased in terms of the
heterotic string, though we later construct various type IIB and
M-theory duals. The vacua are typically non-geometric in the sense
that they are locally geometric, satisfying the supergravity
equations of motion, but globally well-defined only in string
theory. In particular, the complexified K\"ahler modulus will undergo
non-trivial monodromies sourced by assorted heterotic $5$-branes.
We construct some simple
examples and describe how to build general compactifications of this
type.

In section~\ref{hettorsion}, we construct new non-geometric
heterotic solutions with more general torsion. Such spaces have
metrics which are locally non-K\"ahler.  We do this by dualizing
certain M-theory compactifications with flux which played a
prominent role in constructing the first torsional (geometric)
backgrounds~\cite{Dasgupta:1999ss}. The local supersymmetry
constraints on the metrics and fluxes for these kinds of backgrounds
were explored in~\cite{Becker:2009df}.

These heterotic solutions, in turn, also have dual type IIB and
M-theory descriptions, obtained in section~\ref{iibm}, that exhibit
novel characteristics. These are the compact U-folds sought
in~\cite{Kumar:1996zx}\ but of a quite different local form.  In
particular, the space-time supersymmetry spinors have a more general
structure than is usually considered. This allows us to construct,
for example, four-dimensional type IIB compactifications with
three-form flux that is not necessarily of $(2,1)$ Hodge type. We
give an explicit example of such a construction and describe its
M-theory lift.

\vskip 0.5cm \noindent {\bf Note added:} We should  mention that the
solutions found in sections~\ref{hettorsion}\ and~\ref{iibm}\ were
obtained quite some time ago. During the completion of the project,
several papers appeared with interesting related
observations~\cite{Cvetic:2007ju, Koerber:2007hd,Marchesano:2007vw,
Andriot:2008va, Schulgin:2008fv, Vegh:2008jn, ReidEdwards:2008rd}.
It is also worth mentioning a very recent interesting conjecture
that the interpretation of black hole entropy might require the use
of exotic branes associated to non-geometric
monodromies~\cite{deBoer:2010ud}.


\section*{Acknowledgements}
It  is our pleasure to thank Vijay Kumar, Andreas Malmendier,
Ilarion Melnikov, and Wati Taylor for helpful discussions. S.~S.
would also like to thank the Aspen Center for Physics, the KITP
Santa Barbara, and the University of Amsterdam for hospitality
during the completion of this project.   J.~M. would like to thank
TASI 2007, The Simons Workshop in Mathematical Physics 2007, 2008,
2009, the KITP Santa Barbara, and the Mathematical Institute,
University of Oxford for their hospitality during the completion of
this project.

\vskip 0.1 in \noindent  J.~M. is supported in part by an EPSRC Postdoctoral Fellowship EP/G051054/1.
D.~R.~M. is supported in part by National Science Foundation Grant
No.~DMS-0606578.
S.~S. is supported in part by
NSF Grant No.~PHY-0758029, NSF Grant No.~0529954 and the Van der
Waals Foundation.
Any opinions, findings, and conclusions or recommendations expressed in this
material are those of the authors
and do not necessarily reflect the views of the granting agencies.

\section{F-theory and the heterotic string revisited} \label{f-het}

\subsection{$\operatorname{SL}(2,\mathbb{Z})$-invariant scalar fields}
\label{sectionscalarfields}

Following our introductory comments, let us consider a physical
theory which contains a scalar field $\tau$
invariant under an
$\operatorname{SL}(2,\mathbb{Z})$ action. It is natural to try to
construct compactifications of this theory which exploit the
$\operatorname{SL}(2,\mathbb{Z})$-invariance of the scalar.
The general framework for doing so was laid out in~\cite{MR1059826}\
in the language of cosmic strings: the compactification space
should have a multi-valued function $\tau$ on it, defined away from
certain defects of codimension two,
which will undergo $\operatorname{SL}(2,\mathbb{Z})$
transformations around loops encircling the defects. These defects are depicted in figure~\ref{fibration}.

The general problem of specifying such a multi-valued function arose
in the work of Kodaira on elliptically fibered complex manifolds
more than 45 years ago~\cite{MR0184257}.  Any such
elliptically fibered manifold gives rise to a multi-valued function $\tau$
defined on the base of the family, away from the subset of the base at which
singular fibers are located.  Conversely, given the multi-valued
function $\tau$, one can construct in a natural way an elliptically fibered manifold with
fibers $\mathbb{C}/(\mathbb{Z}\oplus\mathbb{Z}\tau)$
over this subset of the base,
which has the additional property that the family
has a section (corresponding to $0\in\mathbb{C}$).\footnote{This result
was obtained by Kodaira \cite{MR0184257} when the base has complex
dimension one, and subsequently generalized by Kawai \cite{MR0216520}
to dimension two and by Ueno \cite{MR0360582} to arbitrary dimension.}

To close this circle of ideas, Kodaira showed that one can pass from
an arbitrary elliptically fibered manifold to its associated
``Jacobian fibration'' (the one with the same $\tau$ function, and a
section) in a natural way that
does not involve finding $\tau$ explicitly.\footnote{This is closely
related to finding an equation in Weierstrass form, as described in
an algebraic context by Deligne \cite{MR0387292}, and explored in this
geometric context
by Nakayama \cite{MR977771}.} Moreover, Kodaira gave
a way to characterize the set of all elliptically fibered manifolds
with a fixed Jacobian fibration when the base has complex dimension
one. This was
later extended to bases of higher complex
dimension by Nakayama \cite{MR1929795,MR1918053}.

As Kodaira explained, two pieces of data are needed to specify $\tau$:
the natural $\operatorname{SL}(2,\mathbb{Z})$-invariant
function $j=j(\tau)$ on the base (which Kodaira called the
``functional invariant'') and the precise
$\operatorname{SL}(2,\mathbb{Z})$ action
on $\tau$, which can be equivalently thought of as the
varying family of integer homology
groups $H_1(\mathbb{C}/(\mathbb{Z}\oplus\mathbb{Z}\tau),\mathbb{Z})$
over the base (which Kodaira called the ``homological invariant'').

Given an elliptically fibered manifold $Z\to S$ with a section, there
is a description of $S$ as a {\em Weierstrass model}\/ (cf.\ \cite{MR0387292}).
That is, there is a $\mathbb{P}^2$-bundle over $S$, and a birational
map from $Z$
to this $\mathbb{P}^2$-bundle, whose image has an (affine)
equation of the form\footnote{When comparing with \cite{MR0387292},
one should bear in mind that we are working over the complex numbers,
so the exceptions to this form having to do with fields of
characteristic $2$ or $3$ do not apply.}
\be\label{weierform} y^2 = x^3 + f(s) x + g(s), \ee
where $f(s)$ and and $g(s)$ are sections of appropriate line bundles over $S$.
To be precise, there is a line bundle
$\mathcal{O}(L)$ on $S$
such that
$f(s)\in H^0(\mathcal{O}(4L))$, $g(s)\in H^0(\mathcal{O}(6L))$;
 we can regard $x$ as a local section of
$\mathcal{O}(2L)$
 and $y$
as a local section of
$\mathcal{O}(3L)$
 with the $\mathbb{P}^2$-bundle
described as
\begin{equation}
\mathbb{P}\left(\mathcal{O}\oplus \mathcal{O}(2L)
\oplus \mathcal{O}(3L)\right).
\end{equation}
The total space may be singular, since certain subvarieties
may be blown down in passing
from the original elliptic fibration to the Weierstrass model.

The fibers of the Weierstrass model
are singular\footnote{Note that a singular point of a fiber is not
necessarily a singular point of the total space, but for every singular
point of the total space, the fiber passing through that point is singular.}
 at the zeroes of the {\em discriminant}\/
\be\label{discrim} \Delta(s) = 4 f(s)^3 + 27 g(s)^2,\ee
and the functional invariant (the $j$-function) is given by the formula
\begin{equation} \label{eq:j}
 j(s)=1728 \frac{4f(s)^3}{4 f(s)^3 + 27 g(s)^2}.
\end{equation}
We will later make use of an equivalent formula for $j(s)-1728$:
\begin{equation} \label{eq:j-1728}
j(s)-1728 = -1728 \frac{27 g(s)^2}{4 f(s)^3 + 27 g(s)^2}.
\end{equation}

The homological invariant is determined by Kodaira's famous table,
reproduced as Table~\ref{tab:kodaira}.  In that table, along any divisor $D$
within $S$ one calculates the orders of vanishing of $f(s)$, $g(s)$ and
$\Delta(s)$ along $D$ and learns about the singularity of
the Weierstrass model over a general point of $D$, as well as the
conjugacy class of the monodromy transformation
about a loop encircling $D$.  It is the latter which determines
the homological invariant.

The last line of the table indicates a
``non-minimal''
Weierstrass equation: one whose singularities
can be improved by making a birational transformation
\begin{equation}
(x,y)\, \mapsto \, \left( x/\psi(s)^2,y/\psi(s)^3 \right),
\end{equation}
(together
with replacing
$\mathcal{O}(L)$ by $\mathcal{O}(L+D)$), where $\psi(s)$ is a section of $\mathcal{O}(D)$ vanishing along $D$.
This birational transformation does not affect the elliptic fibration
away from the singular fibers
in any way, and after a finite number of such improvements, a ``minimal''
Weierstrass model is obtained (that is, one which fits into one of
the earlier lines of the table).  Because each non-minimal Weierstrass
equation can be reduced to a minimal one by this process, it is customary
to focus on the ``minimal'' case.
We will comment below on an additional reason that non-minimal Weierstrass
equations would be unsuitable for the physical applications we have in mind.

Note that the Weierstrass equation is not uniquely specified by the $\tau$ function:
we are free to rescale
\begin{equation}
(x,y,f,g) \mapsto \left(u(s)^{2}x,u(s)^{3}y,u(s)^4f(s),u(s)^6g(s)\right),
\end{equation}
using a nowhere vanishing function
 $u(s)$; this must be taken into account when describing the parameters
of this construction.\footnote{Note that allowing $u(s)$ to be
a section of a line bundle would provide no greater generality,
since a nowhere-vanishing section would trivialize the line bundle.}

\begin{table}
{\footnotesize
\begin{center}
\begin{tabular}{|c|c|c|c|c|c|} \hline
&$\operatorname{ord}_D(f)$&$\operatorname{ord}_D(g)$
&$\operatorname{ord}_D(\Delta)$
&singularity&monodromy\\ \hline\hline
$I_0$&$\ge0$&$\ge0$&$0$&none
&$\begin{pmatrix}\hphantom{-}1&\hphantom{-}0\\\hphantom{-}0&\hphantom{-}1\end{pmatrix}$\\ \hline
$I_n$, $n\ge1$&$0$&$0$&$n$&$A_{n-1}$
&$\begin{pmatrix}\hphantom{-}1&\hphantom{-}n\\\hphantom{-}0&\hphantom{-}1\end{pmatrix}$\\ \hline
$II$&$\ge1 $&$   1  $&$    2 $&  none
&$\begin{pmatrix}\hphantom{-}1&\hphantom{-}1\\-1&\hphantom{-}0\end{pmatrix}$\\ \hline
$III$&$  1 $&$   \ge2 $&$   3 $&$  A_1$
&$\begin{pmatrix}\hphantom{-}0&\hphantom{-}1\\-1&\hphantom{-}0\end{pmatrix}$\\ \hline
$IV$&$ \ge2 $&$  2  $&$    4 $&$  A_2$
&$\begin{pmatrix}\hphantom{-}0&\hphantom{-}1\\-1&-1\end{pmatrix}$\\ \hline
$I_0^*$&$\ge2$&$\ge3$&$6$&$D_{4}$
&$\begin{pmatrix}-1&\hphantom{-}0\\\hphantom{-}0&-1\end{pmatrix}$\\ \hline
$I_n^*$, $n\ge1$&$2$&$3$&$n+6$&$D_{n+4}$
&$\begin{pmatrix}-1&\hphantom{-}n\\\hphantom{-}0&-1\end{pmatrix}$\\ \hline
$IV^*$&$\ge3$&$  4$  &$  8$&$   E_6$
&$\begin{pmatrix}-1&-1\\\hphantom{-}1&\hphantom{-}0\end{pmatrix}$\\ \hline
$III^*$&$  3 $&$   \ge5 $&$   9 $&$  E_7$
&$\begin{pmatrix}\hphantom{-}0&-1\\\hphantom{-}1&\hphantom{-}0\end{pmatrix}$\\ \hline
$II^*$&$ \ge4$&$   5   $&$   10 $&$  E_8$
&$\begin{pmatrix}\hphantom{-}0&-1\\\hphantom{-}1&\hphantom{-}1\end{pmatrix}$\\ \hline
non-minimal&$\ge4$&$\ge6$&$\ge12$&non-canonical&--\\ \hline
\end{tabular}
\end{center}
\medskip
\caption{Kodaira's classification of singular fibers and monodromy}\label{tab:kodaira}
}
\end{table}

Kodaira also gave a formula for the canonical bundle of the total space
of a minimal Weierstrass fibration when the base has complex dimension one
(subsequently extended by others to higher dimension
under certain hypotheses).  The formula states
that
\begin{equation}\label{eq:canon}
 \mathcal O(12 K_{\overline{Z}}) =
\pi^*(\mathcal O(12 K_S + \Delta)),
\end{equation}
where $\pi:\overline{Z}\to S$ is the Weierstrass fibration.

To summarize: the data of a locally defined
$\operatorname{SL}(2,\mathbb{Z})$-invariant scalar $\tau$ on some
manifold $S$ can be given in terms of an elliptic fibration $Z\to S$
with a section, and is effectively given by specifying a line bundle
$\mathcal{O}(L)$
 and describing $Z$ as the desingularization of a hypersurface
$\overline{Z}$ in the $\mathbb{P}^2$-bundle
\begin{equation}
\mathbb{P}\left(\mathcal{O}\oplus\mathcal{O}(2L)
\oplus\mathcal{O}(3L)\right)\to S,
\end{equation}
defined by a Weierstrass equation
\begin{equation}\label{eq:weierstrass}
 y^2=x^3+f(s)x+g(s),
\end{equation}
which does not fall into the last line of Table~\ref{tab:kodaira}
for any divisor $D$ on $S$.

\subsection{F-theory}
\label{ftheory}

The F-theory construction  is a familiar
application of the discussion in the
previous section~\cite{Vafa:1996xn,FCY1,FCY2}.  F-theory is a description of general
type IIB string backgrounds in which the complexified string coupling $\tau_F$ of the theory
is allowed to be multi-valued and is defined
away from defects of codimension two.

Kodaira's table allows a classification of the defects, using monodromy:
a stack of $n$ D7-branes
corresponds to Kodaira's type $I_n$; a stack of
 $n$ D7-branes on top of an orientifold
O7-plane corresponds to Kodaira's type $I_n^*$; and various exotic $7$-branes
which are difficult to analyze from a perturbative string perspective
correspond
to the remaining Kodaira types $II$, $III$, $IV$, $IV^*$, $III^*$, $II^*$.

There are special cases of the F-theory construction in which the $\tau$
function is constant \cite{Sen:1996vd,Dasgupta:1996ij}.  First, for
any constant value of the F-theory function $\tau_F$ we can choose data
of the form
\be
f(s) = \varphi h(s)^2, \quad g(s) = \gamma h(s)^3,
\ee
for some section $h(s)$ of the line bundle $\mathcal{O}(2L)$,
and constants $\varphi$ and $\gamma$.
In this case,
\be
 j(s) = 1728\cdot \frac{4\varphi^3}{4\varphi^3+27\gamma^2} = j(\tau_F),
\ee
is the constant value.  The singular fibers occur at the zeros of $h(s)$,
and are all of
Kodaira type $I_0^*$, which corresponds to $\SO(8)$ enhanced gauge symmetry.
(If the locus $h(s)=0$ is reducible, there can be more than one $\SO(8)$
component.)  This construction is equivalent to one made with
orientifold planes and can be
studied perturbatively (cf.~\cite{Sen:1997gv,Sen:1996vd}) by choosing
$\tau_F$ near $i\infty$.

Secondly, if we take $f$ to be identically zero, then we end up with
$\tau_F=e^{2\pi i/3}$ while thirdly, if we take $g$ to be identically zero,
then we find $\tau_F=i$.  Various Kodaira fibers and enhanced gauge
symmetry groups are possible in these cases.  Since $\tau_F$ is fixed away
from $i\infty$ in these cases, a purely perturbative analysis is not possible.

Our confidence in F-theory is bolstered by
F-theory/M-theory duality: after compactifying F-theory on an additional
circle, one finds an equivalence with M-theory compactified on the
elliptically fibered manifold $Z$, or more precisely, on the total
space $\overline{Z}$ of
the Weierstrass fibration.\footnote{This total space may
have singularities, as indicated in Table~\ref{tab:kodaira},
and such singularities in an M-theory compactification give rise
to non-abelian gauge symmetries of the compactified theory~\cite{Aspinwall:1995xy,enhanced,klemmmayr,WitMF}.
A non-minimal Weierstrass fibration will have a singularity which
is non-canonical, that is, which does not preserve the holomorphic
form of top degree on the fibration, and for this reason, such
fibrations are not generally allowed when studying compactifications
of M-theory or F-theory.}
Thus, to get a supersymmetric compactification of F-theory, we require
$\overline{Z}$ to be Calabi--Yau,
which---thanks to eq.~\eqref{eq:canon}---happens when
$\mathcal O(12K_S+\Delta)$ is trivial.  Since $\mathcal O(\Delta)=
\mathcal O(12L)$, we should choose $\mathcal O(L)=\mathcal O(-K_S)$
(possibly up to torsion) to ensure that $\overline{Z}$ is  Calabi--Yau
(with at most canonical singularities).

In section~\ref{constrnongeom}, we will
construct some new non-geometric compactifications of the heterotic
strings, and will make use of a similar confidence-building
duality: the corresponding F-theory/heterotic duality.  In
section~\ref{constrnongeom},
we explain how those F-theory/heterotic
dualities---in the absence of Wilson lines---are much
more geometric than had originally been realized.  The key insight
about those dualities was found by Clingher and Doran~\cite{math.AG/0602146},
based in part on some old work of the second author of this paper~\cite{k3Picard};
our discussion is based on a refinement of these ideas.

First, though, we need to analyze F-theory models with certain large
gauge groups.
In anticipation of the duality with the heterotic string (to be
reviewed in the next section),
we construct F-theory models in dimension $8$ and below with gauge groups
 $G=(E_8\times E_8)\rtimes \mathbb Z_2$ or
$G=\operatorname{Spin}(32)/\mathbb{Z}_2$.
In $8$ dimensions, this amounts to giving
an elliptic fibration $Z_G\to\mathbb{P}^1$
with gauge symmetry group $G$.

The Weierstrass model for
$Z_{(E_8\times E_8)\times \mathbb Z_2}$
 was essentially given in
\cite{FCY2} (see also \cite{instK3}): there must be two fibers of
Kodaira type $II^*$.
By choosing an appropriate coordinate $\sigma$ on the base $\mathbb P^1$,
we can assume that these fibers
are located at $\sigma=0$ and $\sigma=\infty$;
the equation then takes the form
\begin{equation}\label{eq:MV} Y^2= X^3+a \sigma^4 X + b \sigma^5 + c \sigma^6  + d \sigma^7,\end{equation}
for some constants $a$, $b$, $c$, $d$.
We review the argument for this in Appendix~\ref{app:Weierstrass}.
Note that the discriminant of eq.~\eqref{eq:MV} is
\be
 \Delta=\sigma^{10}\left(4a^3\sigma^{2}+27\left(b+c\sigma+d\sigma^2
\right)^2\right);
\ee
since the (affine) degree of the discriminant in $\sigma$ is $14$,
there is an implicit
zero of order $10$ at $\sigma=\infty$, the location of the second fiber of
type $II^*$.  To prevent the zeros at $\sigma=0$ and $\sigma=\infty$
from having order
greater than $10$ (which would lead to a non-minimal Weierstrass model),
we should assume that neither $b$ nor $d$
is zero.

To obtain the Weierstrass model for
$Z_{\operatorname{Spin}(32)/\mathbb Z_2}$, we need a fiber of type $I_{12}^*$ and a
Mordell--Weil group of $\mathbb Z_2$ (see
\cite{MR1416960,Aspinwall:1996vc}).
Note that by choosing an appropriate coordinate $s$ on the base,
we can assume that the fiber of type
$I_{12}^*$ is located at $s=\infty$.
In this case, rather than using the traditional Weierstrass equation,
we change coordinates
so that the point of order $2$ on the elliptic curves (which corresponds
to the $\mathbb Z_2$ factor in the Mordell--Weil group) is at $x=0$.
Then, as we review in Appendix~\ref{app:Weierstrass}, the equation takes the form
\begin{equation}\label{eq:AMbis}
 y^2=x^3+(p_0s^3+p_1s^2+p_2s+p_3)x^2+\varepsilon x,
\end{equation}
with discriminant
\be
\Delta=-\varepsilon^2 (p(s)^2-4\varepsilon),
\ee
where
\be
 p(s)=p_0s^3+p_1s^2+p_2s+p_3.
\ee
To ensure that the gauge group is precisely $\operatorname{Spin}(32)/\mathbb Z_2$, we must assume that
neither $\varepsilon$ nor $p_0$ is zero.

Remarkably, these two elliptically fibered $K3$ surfaces
$Z_{(E_8\times E_8)\rtimes\mathbb Z_2}$ and
$Z_{\operatorname{Spin}(32)/\mathbb Z_2}$
are birational to
each other if the coefficients are identified properly;
we will make use of these birational equivalences in our constructions
in the next section.
If we start with the Weierstrass model
$\overline{Z}_{(E_8\times E_8)\rtimes\mathbb{Z}_2}$
given by eq.~\eqref{eq:MV} with $d\ne0$, we can make
a birational change to get to another $K3$ surface:
let $X=x^2s/d^2$, $Y=x^2y/d^3$, $\sigma=x/d$, and multiply the equation
by $d^6/x^4$, to obtain
\be
 y^2 = x^2s^3+ax^2s+bdx+cx^2+x^3.
\ee
This has the form of eq.~\eqref{eq:AMbis} with
\begin{equation}\label{eq:bir1}
p(s)=s^3+as+c \quad \text{and} \quad \varepsilon=bd.
\end{equation}

Conversely,
if we start with the Weierstrass model
 $\overline{Z}_{\operatorname{Spin}(32)/\mathbb{Z}_2}$
described by
eq.~\eqref{eq:AMbis}
and assume $p_0\ne0$,
setting $x=\sigma$, $y=Y/p_0\sigma^2$, $s=\widehat{X}/p_0\sigma^2$
and multiplying by $p_0^2\sigma^4$
we find
\be
 {Y}^2=p_0^2{\sigma}^7+\widehat{X}^3+p_1\sigma^2\widehat{X}^2+p_0p_2\sigma^4\widehat{X}+p_0^2p_3{\sigma}^6+
p_0^2\varepsilon {\sigma}^5.
\ee
To put this into Weierstrass form we need one more change of variables,
completing the cube via $\widehat{X}=X-\frac13p_1\sigma^2$:
\be
 {Y}^2=
X^3
+\left(p_0p_2-\frac13p_1^2 \right)\sigma^4X
+p_0^2\varepsilon {\sigma}^5
+\left(\frac2{27}p_1^3
-\frac13p_0p_1p_2
+p_0^2p_3 \right)\sigma^6
+p_0^2\sigma^7 .
\ee
This has the form of eq.~\eqref{eq:MV} with
\begin{equation}\label{eq:bir2}
\begin{aligned}
a &= p_0p_2-\frac13p_1^2,\\
b&= p_0^2\varepsilon,\\
c&= \frac2{27}p_1^3
-\frac13p_0p_1p_2
+p_0^2p_3,\\
d&=p_0^2.
\end{aligned}
\end{equation}

The existence of these birational isomorphims between the
Weierstrass models $\overline{Z}_{(E_8\times E_8)\rtimes\mathbb Z_2}$
and $\overline{Z}_{\operatorname{Spin}(32)/\mathbb Z_2}$ implies
that the corresponding
 nonsingular surfaces $Z_{(E_8\times E_8)\rtimes\mathbb Z_2}$
and $Z_{\operatorname{Spin}(32)/\mathbb Z_2}$ are isomorphic; however,
the isomorphism does not preserve the elliptic fibrations.  Thus,
if M-theory is compactified on either of these
nonsingular surfaces, the resulting seven-dimensional
theory will have two distinct F-theory limits, corresponding to these
two different elliptic fibrations (with section) on the surface.

For both gauge groups $G$, we can extend the above construction
to a broader
class of F-theory models by considering F-theory on a base $S$ which
is a $\mathbb P^1$-bundle over some space $B$.  We can express $S$
in the form $\mathbb P(\mathcal O\oplus\mathcal O(\Lambda_G))$ for some
line bundle $\mathcal O(\Lambda_G)$ on $B$, with projection map
$\varphi:S\to B$, and regard $\sigma$ and $s$ as
sections of the appropriate
$\mathcal O(\Lambda_G)$.  If $\Sigma_0\subset S$ is the divisor where
$\sigma=0$ in the first case (or $s=0$ in the second case),
and $\Sigma_\infty\subset S$ is the divisor where
$\sigma=\infty$ in the first case (or $s=\infty$ in the second case),
then $\mathcal O(\Sigma_\infty - \Sigma_0)=\varphi^*\mathcal O(\Lambda_G)$ and we
can write
\be
\begin{aligned} \mathcal{O}(-K_S)
&=\mathcal{O}(\Sigma_0+\Sigma_\infty +\varphi^*(-K_B)) \\
&=\mathcal{O}(2\Sigma_0 +\varphi^*(-K_B+\Lambda_G)).
\end{aligned}
\ee
This is the line bundle which we use to build an F-theory model
whose Weierstrass fibration $\overline{Z}$ is Calabi--Yau.

\begin{table}
\medskip
\begin{center}
\begin{tabular}{|l|l|} \hline
$a\sigma^4$ & $\mathcal O(-4K_S) =
\mathcal O\left(4\Sigma_0+4\Sigma_\infty+\varphi^*(-4K_B)\right)$ \\ \hline
$b\sigma^5$ & $\mathcal O(-6K_S) =
\mathcal O\left(5\Sigma_0+7\Sigma_\infty+\varphi^*(-6K_B+\Lambda_{(E_8\times E_8)\rtimes \mathbb Z_2})\right)$ \\ \hline
$c\sigma^6$ & $\mathcal O(-6K_S) =
\mathcal O\left(6\Sigma_0+6\Sigma_\infty+\varphi^*(-6K_B)\right)$ \\ \hline
$d\sigma^7$ & $\mathcal O(-6K_S) =
\mathcal O\left(7\Sigma_0+5\Sigma_\infty+\varphi^*(-6K_B-\Lambda_{(E_8\times E_8)\rtimes \mathbb Z_2})\right)$ \\ \hline
\end{tabular}
\medskip
\caption{The transformation properties of the coefficients in~\C{eq:MV}.}\label{tab:coefficients}
\end{center}
\end{table}

In the case $G=(E_8\times E_8)\rtimes \mathbb Z_2$, we get a Weierstrass
equation of the form eq.~\eqref{eq:MV}.  To determine how the various
coefficients in that equation transform, we illustrate in
Table~\ref{tab:coefficients}
various forms of the appropriate line bundles.
It follows that $a$, $b$, $c$, $d$ are sections of
\begin{equation}
\mathcal O(-4K_B), \,\, \mathcal O(-6K_B+\Lambda_{(E_8\times E_8)\rtimes \mathbb Z_2}),  \,\, \mathcal O(-6K_B),  \,\,
\mathcal O(-6K_B-\Lambda_{(E_8\times E_8)\rtimes \mathbb Z_2}),
\end{equation}
respectively.

\begin{table}
\begin{center}
\begin{tabular}{|l|l|} \hline
$p_0s^3$ & $\mathcal O(-2K_S) =
\mathcal O\left(3\Sigma_0+\Sigma_\infty+\varphi^*(-2K_B-\Lambda_{\operatorname{Spin}(32)/\mathbb Z_2})\right)$ \\ \hline
$p_1s^2$ & $\mathcal O(-2K_S) =
\mathcal O\left(2\Sigma_0+2\Sigma_\infty+\varphi^*(-2K_B)\right)$ \\ \hline
$p_2s$ & $\mathcal O(-2K_S) =
\mathcal O\left(\Sigma_0+3\Sigma_\infty+\varphi^*(-2K_B+\Lambda_{\operatorname{Spin}(32)/\mathbb Z_2})\right)$ \\ \hline
$p_3$ & $\mathcal O(-2K_S) =
\mathcal O\left(4\Sigma_\infty+\varphi^*(-2K_B+2\Lambda_{\operatorname{Spin}(32)/\mathbb Z_2})\right)$ \\ \hline
$\varepsilon$ & $\mathcal O(-4K_S) =
\mathcal O\left(8\Sigma_\infty+\varphi^*(-4K_B+4\Lambda_{\operatorname{Spin}(32)/\mathbb Z_2})\right)$ \\ \hline
\end{tabular}
\caption{The transformation properties of the coefficients in~\C{eq:AMbis}.}\label{tab:coeff}
\end{center}
\end{table}

Similarly, in the case of $G=\operatorname{Spin}(32)/\mathbb Z_2$,
we get a Weierstrass equation of the form eq.~\eqref{eq:AMbis}, whose
coefficients are analyzed in Table~\ref{tab:coeff}.
It follows that $(p_0, p_1, p_2, p_3, \varepsilon)$ are sections of
\begin{equation}
\mathcal O(-2K_B-\Lambda_{\operatorname{Spin}(32)/\mathbb Z_2}), \,\, \mathcal O(-2K_B), \,\, \mathcal O(-2K_B+\Lambda_{\operatorname{Spin}(32)/\mathbb Z_2}), \,\,
\mathcal O(-2K_B+2\Lambda_{\operatorname{Spin}(32)/\mathbb Z_2}),
\end{equation}
and $ \mathcal O(-4K_B+4\Lambda_{\operatorname{Spin}(32)/\mathbb Z_2})$, respectively.

Notice that the birational equivalence between the two models also extends
to this higher-dimensional context, once we identify the line bundles
correctly.  Starting from $G=(E_8\times E_8)\rtimes\mathbb Z_2$ using
an arbitrary line bundle
$\mathcal{O}(\Lambda_{(E_8\times E_8)\rtimes\mathbb Z_2})$,
we get a dual model with line bundle
\be
 \mathcal{O}(\Lambda_{\operatorname{Spin}{32}/\mathbb Z_2}) =
\mathcal{O}(-2K_B),
\ee
compatible with eq.~\eqref{eq:bir1}.
Conversely, starting from $G=\operatorname{Spin}{32}/\mathbb Z_2$
and an arbitrary line bundle
$\mathcal{O}(\Lambda_{\operatorname{Spin}{32}/\mathbb Z_2})$,
we get a dual model with line bundle
\be
 \mathcal{O}(\Lambda_{(E_8\times E_8)\rtimes\mathbb Z_2}) =
\mathcal{O}(-2K_B+
2\Lambda_{\operatorname{Spin}{32}/\mathbb Z_2}),
\ee
compatible with eq.~\eqref{eq:bir2}.

\subsection{F-theory/heterotic dualities}
\label{fhetdualities}

The duality between F-theory and the heterotic string in
dimension $8$, originally
proposed by Vafa~\cite{Vafa:1996xn}, takes the following form when
the heterotic gauge group is unbroken: for heterotic gauge group $G$,
there is a family of elliptically fibered $K3$ surfaces $(X_G)_z$ (with section)
parameterized by,
\be
z\in SO(2,2;\mathbb Z)\backslash SO(2,2)/SO(2)\times SO(2),
\ee
and a family of heterotic string vacua $(Y_G)_z$ with gauge group $G$,
such that F-theory on $(X_G)_z$ is dual to the heterotic string vacuum
$(Y_G)_z$.

The data needed to specify the heterotic vacuum $(Y_G)_z$ consists
of a flat metric and a $\B$-field on a two-torus.\footnote{Since $G$ is
unbroken, all Wilson line expectation values must vanish.}  There is a unique
complex structure compatible with any given metric, so this data
can be expressed as an elliptic curve $E$ (i.e., a two-torus equipped
with complex structure), as well as a K\"ahler class and $\B$-field on $E$.
These latter two can be combined into the complex number $\rho$, defined in~\C{definerho},
which naturally lives in the upper half-plane and is invariant under the
$SL(2,\mathbb Z)$ action.  Similarly, the complex structure
on $E$ can be represented by a complex number $\tau$ in the upper half-plane, modulo $SL(2,\mathbb Z)$.
The duality between F-theory and the heterotic string suggests that
for each F-theory vacuum with gauge group $G$,
$\tau$ and $\rho$ should be expressible as functions of the coefficients
of \eqref{eq:MV} or \eqref{eq:AMbis}.

One should note that much of the discussion in the literature, including
the analysis in \cite{FCY1,FCY2}, is limited to a particular limit,
in which $({a^3}/{bd})\to\infty$ and $({c^2}/{bd})\to\infty$ while
$({c^2}/{a^3})$ remains finite.\footnote{Of
course, there are instances where two different limits of this kind
are taken in order to study a duality. This was done
for example in~\cite{FCY1}\
which studied the duality of~\cite{DuffMinasianWitten}.}
As we will explain shortly, from the heterotic point of view this
is equivalent to taking the large volume limit
$\rho\rightarrow i\infty$, where the heterotic
supergravity description is good. From the point of view of F-theory
there is nothing special about this limit. One could consider
generic values of $({a^3}/{bd})$ and $({c^2}/{bd})$ in $\CC$,
 in which case the heterotic torus $T^2$
has some finite size and complex structure. As we will see, the fibered
version of this case corresponds to non-geometric heterotic compactifications.

In fact, the explicit correspondence between F-theory and heterotic
parameters in 8 dimensions was calculated in the case of
$G=(E_8\times E_8)\rtimes\mathbb Z_2$ in the early days of F-theory
\cite{LopesCardoso:1996hq, Lerche:1998nx}.  In the notation of the present
paper\footnote{To compare the two, one must make the substitution
\be X=b^{7/6}d^{-5/6}\widetilde{X}, Y=b^{7/4}d^{-7/4}\widetilde{Y},
\sigma=b^{1/2}d^{-1/2}\widetilde{\sigma}, \ee in eq.~\eqref{eq:MV}
and multiply by $d^{5/2}b^{-7/2}$ to obtain \be
 \widetilde{Y}^2 = \widetilde{X}^3 + ab^{-1/3}d^{-1/3}\widetilde{\sigma}^4
\widetilde{X} + \widetilde{\sigma}^5 +
cb^{-1/2}d^{-1/2}\widetilde{\sigma}^6 + \widetilde{\sigma}^7. \ee },
the authors of \cite{LopesCardoso:1996hq, Lerche:1998nx}\
found:\footnote{These same formulas were independently discovered in the
mathematics literature in a slightly different context \cite{MR2279280}.}
\begin{equation}\label{eq:taurho1}
j(\tau)j(\rho) = -1728^2 \frac{a^3}{27bd},
\end{equation}
\begin{equation}\label{eq:taurho2}
\left(j(\tau)-1728 \right) \left(j(\rho)-1728 \right) = 1728^2\frac{c^2}{4bd}.
\end{equation}
which implies that
\begin{equation}\label{eq:taurho3}
\frac{c^2}{a^3} = - \left(1-\frac{1728}{j(\tau)}\right)\left(1-\frac{1728}{j(\rho)}\right);
\end{equation}
the large volume heterotic limit $j(\rho)\to\infty$ thus
corresponds to $({a^3}/{bd})\to\infty$ and $({c^2}/{bd})\to\infty$ while
$({c^2}/{a^3})$ remains finite.

Analogous formulae were found much more recently
\cite{math.AG/0602146} for the case
$G=\operatorname{Spin}{32}/\mathbb{Z}_2$. Our goal in this
subsection is to refine these formulae in both cases, and to give a
much more geometric explanation of them.

As stressed in section~\ref{sectionscalarfields}, the heterotic
elliptic curve $E$ naturally
encodes the information provided by the modular function $\tau$.  Similarly,
since $\rho$ is also an $SL(2,\mathbb Z)$ modular function, we can
encode the information it provides in a second elliptic curve $F$.
In dimension $8$ this is not so crucial, but when we go to lower dimension,
and want to use $\tau$ and $\rho$ as fields which can vary in the
compactification to lower dimension (exploiting the $SL(2,\mathbb Z)$
symmetry), this is an important step.

As Clingher and Doran showed \cite{math.AG/0602146},
the geometric connection between the heterotic and F-theory sides of
this story is provided (in the absence of Wilson lines)
by the notion of a Shioda--Inose
structure for $K3$ surfaces.  Following \cite{k3Picard},  we say that
a $K3$ surface $Z$ has a {\em Shioda--Inose structure}\/ if there is an
automorphism $\iota:Z\to Z$ of order two, preserving the holomorphic
$2$-form, and a complex torus $A$ of complex dimension $2$, such
that $Z/\iota$ is birationally isomorphic to the Kummer surface
$A/(-1)$.  This definition was motivated by work of Shioda and Inose
 who considered such structures in special cases~\cite{Shioda-Inose,MR578868}.

The main theorem of \cite{k3Picard} (combined with some known facts
about the N\'eron--Severi group of a complex torus
\cite{Mumford:av,kuga-satake})
implies that the $K3$ surfaces $Z_G$
constructed in section~\ref{ftheory}\ have Shioda--Inose structures
with the complex torus taking the form $E\times F$ for
two elliptic curves $E$ and $F$.  This is the geometric form
of F-theory/heterotic duality: the elliptic curves $E$ and $F$
associated to $Z_G$ provide the data for the heterotic vacuum.

Clingher and Doran \cite{math.AG/0602146}
have constructed the Shioda--Inose structure for
$Z_{\operatorname{Spin}{32}/\mathbb{Z}_2}$
in a very explicit
manner, and we refine their result in Appendix~\ref{CDrefinement}.  The result is stated
in the opposite direction from the discussion above:  starting
with Weierstrass equations
\be
 v^2=u^3+\lambda_2u+\lambda_3, \quad \text{and} \quad
 w^2=z^3+\mu_2z+\mu_3,
\ee
defining two elliptic curves $E$ and $F$,  respectively, the equation for
the associated F-theory (Weierstrass) elliptic fibration
$\overline{Z}_{\operatorname{Spin}{32}/\mathbb{Z}_2}$
is given by
\be \label{eq:spineq}
 y^2 = x^3 + (s^3 -3 \lambda_2 \mu_2 s - \frac{27}2 \lambda_3 \mu_3)x^2
+\frac1{16} (4\lambda_2^3+27\lambda_3^2)(4\mu_2^3+27\mu_3^2)x.
\ee
In fact, letting $\iota_{\operatorname{Spin}{32}/\mathbb{Z}_2}$ be the
automorphism of
$\overline{Z}_{\operatorname{Spin}{32}/\mathbb{Z}_2}$ defined by translation
by the
point of order $2$ in the Mordell--Weil group, the quotient
$Z_{\operatorname{Spin}{32}/\mathbb{Z}_2}/\iota_{\operatorname{Spin}{32}/\mathbb{Z}_2}$ is birationally isomorphic
to the Kummer surface $(E\times F)/(-1)$.
(See Appendix~\ref{CDrefinement} for the details of this.)

From this, and the birational equivalence we found between
$\overline{Z}_{\operatorname{Spin}{32}/\mathbb{Z}_2}$ and
$\overline{Z}_{(E_8\times E_8)\rtimes\mathbb{Z}_2}$,
we can find a model for the
$G=(E_8\times E_8)\rtimes\mathbb{Z}_2$ case as well.  This time, we
need to choose two factorizations
\be
\begin{aligned}
\frac14(4\lambda_2^3+27\lambda_3^2) &= b(\lambda)d(\lambda), \\
\frac14(4\mu_2^3+27\mu_3^2) &= b(\mu) d(\mu),
\end{aligned}
\ee
and then the equation of
$\overline{Z}_{(E_8\times E_8)\rtimes\mathbb Z_2}$ takes the form
\be
 Y^2 = X^3   -3\lambda_2\mu_2 \sigma^4 X + b(\lambda) b(\mu)\sigma^5
- \frac{27}2\lambda_3\mu_3 \sigma^6 + d(\lambda) d(\mu)\sigma^7.
\ee
In this case, the Shioda--Inose structure is induced by the
automorphism $\iota_{(E_8\times E_8)\rtimes\mathbb{Z}_2}$
which acts on the base of the elliptic fibration to exchange the two fibers
of type $II^*$, and acts on the fiber by multiplication by $-1$; it
can be written as
\be
 \iota_{(E_8\times E_8)\rtimes\mathbb{Z}_2}:
(X,Y,\sigma)\mapsto (\frac{b^2X}{d^2\sigma^4},
\frac{-b^3Y}{d^3\sigma^6}, \frac{b}{d \sigma}),
\ee
where $b=b(\lambda)b(\mu)$ and $d=d(\lambda)d(\mu)$.
Once again, the quotient
$Z_{(E_8\times E_8)\rtimes\mathbb Z_2}/\iota_{(E_8\times E_8)\rtimes\mathbb{Z}_2}$ is birationally isomorphic
to the Kummer surface $(E\times F)/(-1)$.

Let us verify that eqs.~\eqref{eq:taurho1} and \eqref{eq:taurho2} are
satisfied for this elliptic fibration.  Since
$bd=b(\lambda)d(\lambda)b(\mu)d(\mu)$ we have
\be
\begin{aligned}
-1728^2\frac{a^3}{27bd} &= -1728^2\frac{(-3\lambda_2\mu_2)^3}{%
\frac{27}{16}(4\lambda_2^3+27\lambda_3^2)(4\mu_2^3+27\mu_3^2)}\\
&= 1728^2 \frac{(4\lambda_2^3)(4\mu_2^3)}{%
(4\lambda_2^3+27\lambda_3^2)(4\mu_2^3+27\mu_3^2)} = j(\tau)j(\rho),
\end{aligned}
\ee
using eq.~\eqref{eq:j}, and
\be
\begin{aligned}
1728^2\frac{c^2}{4bd} &= 1728^2\frac{(-\frac{27}2 \lambda_3\mu_3)^2}{%
\frac{4}{16}(4\lambda_2^3+27\lambda_3^2)(4\mu_2^3+27\mu_3^2)}\\
&=(-1728)^2\frac{(27\lambda_3^2)(27\mu_3^2)}{%
(4\lambda_2^3+27\lambda_3^2)(4\mu_2^3+27\mu_3^2)} \\&= (j(\tau)-1728)(j(\rho)-1728),
\end{aligned}
\ee
using eq.~\eqref{eq:j-1728},
verifying the formulas derived in \cite{LopesCardoso:1996hq}.

\section{Non-geometric heterotic models}
\label{NGHM}

\subsection{Constructing non-geometric heterotic models}
\label{constrnongeom}

In this section, we wish to use the duality we have analyzed to construct
F-theory duals to various heterotic models.  We begin with an elliptically
fibered space $\mathcal E\to B$, and consider the heterotic string on this space with
unbroken gauge group.  Maintaining unbroken gauge group requires two
things: all Wilson lines must be trivial, and all instantons must be
pointlike.\footnote{There is also the possibility of ``hidden obstructors''
which do not break the gauge group~\cite{Aspinwall:1996vc}, but we do not consider those here.}
The complex structure on the total space $\mathcal E$ determines complex structures
on the elliptic fibers, but the complexified K\"ahler class on the fiber
is left undetermined.

For simplicity, we assume that the elliptic fibration
$\mathcal E\to B$ has a section, but in principle our construction can be made
without that requirement.  Under this assumption, $\mathcal E$ can be described
by an equation
\be
 v^2=u^3+\lambda_2u+\lambda_3,
\ee
where $\lambda_2$ and $\lambda_3$ are sections of appropriate line bundles
$\mathcal{O}(4L_\tau)$ and $\mathcal{O}(6L_\tau)$ on $B$.
Note that to completely specify the geometry,
we must also specify the locations of the point-like instantons
on the space $\mathcal E$; we will return to this point later.

To build a (possibly) non-geometric model, we wish to allow the
complexified K\"ahler parameter to be a non-constant
function $\rho$ on the base $B$.
Strictly speaking, there will be some defect locus $\Delta_\rho$ at points
of which either
$\rho$ is multiple-valued or $\rho$ approaches infinity, so that $\rho$
is only  well-defined on $B-\Delta_\rho$.  Moreover, there is an
$\operatorname{SL}(2,\mathbb{Z})$ ambiguity of $\rho$, so even on
$B-\Delta_\rho$, $\rho$ is only locally well-defined.

Hellerman, McGreevy, and Williams \cite{Hellerman:2002ax}
took a ``stringy cosmic string'' point of view \cite{MR1059826}
in specifying the
function $\rho$, but here we do something much closer in spirit to
the construction of
$F$-theory: we specify $\rho$ via an auxiliary elliptic fibration
$\pi_\rho:\mathcal F\to B$, so that the periods of the elliptic curve
$\pi^{-1}(b)$ are $\mathbb{Z}\oplus \mathbb{Z}\rho(b)$.
Just as in $F$-theory, in order to specify $\rho$ in this way, we
can assume that $\pi_\rho:\mathcal F\to B$ has a section.  Thus, $\mathcal F$ will have a
Weierstrass equation:
\be
 w^2=z^3+\mu_2z+\mu_3,
\ee
where $\mu_2$ and $\mu_3$ are sections of appropriate line bundles
${\mathcal O}(4L_\rho)$ and ${\mathcal O}(6L_\rho)$ on the
base $B$.

Because our construction does not necessarily have a large radius
limit where supergravity techniques can be employed, we will derive
certain restrictions on the families $\mathcal E$ and $\mathcal F$
indirectly via duality with  $F$-theory. The restrictions to which
we refer are the analogues of the restriction that the total space
of $\mathcal{E}$ be Calabi--Yau if $\rho$ is constant. In the
$\operatorname{Spin}{32}/\mathbb{Z}_2$ case, the $F$-theory dual is
given by eq.~\eqref{eq:spineq}, where now the coefficients $p_0$,
\dots, $p_3$, $s$ are considered as sections of appropriate line
bundles. Comparing line bundles, we see that \be
\begin{aligned}
\mathcal{O}
  &= \mathcal{O}(-2K_B -\Lambda_{\operatorname{Spin}{32}/\mathbb{Z}_2}),\\
\mathcal{O}(4L_\tau + 4L_\rho)
  &= \mathcal{O}(-2K_B+\Lambda_{\operatorname{Spin}{32}/\mathbb{Z}_2}),\\
\mathcal{O}(6L_\tau + 6L_\rho)
  &= \mathcal{O}(-2K_B+2\Lambda_{\operatorname{Spin}{32}/\mathbb{Z}_2}),\\
\mathcal{O}(12L_\tau + 12L_\rho)
  &= \mathcal{O}(-4K_B + 4\Lambda_{\operatorname{Spin}{32}/\mathbb{Z}_2}),\\
\end{aligned}
\ee
where the first relation comes from the fact that $p_0$ is non-vanishing.
Thus,
$\mathcal{O}(\Lambda_{\operatorname{Spin}{32}/\mathbb{Z}_2})=\mathcal{O}(-2K_B)$
 and $\mathcal{O}(L_\tau+L_\rho)=\mathcal{O}(-K_B)$ (up to torsion).

It follows that for a given base $B$, we will be able to construct a
non-geometric compactification for the $\operatorname{Spin}{32}/\mathbb{Z}_2$
heterotic string out of any two elliptic fibrations $\pi_\tau:\mathcal E\to B$
and $\pi_\rho:\mathcal F\to B$, provided that the associated line bundles
$\mathcal{O}(L_\tau)$ and $\mathcal{O}(L_\rho)$ satisfy
\be
\mathcal{O}(L_\tau+L_\rho)=\mathcal{O}(-K_B),
\ee
up to torsion.

We can also find an F-theory dual in the case of the
$(E_8\times E_8)\rtimes\mathbb{Z}_2$ heterotic string.
For this, we need to specify a factorization of
$\varepsilon$ into $bd$, where $b$ and $d$ are sections of appropriate
line bundles.  Since $\varepsilon$ is itself a product, this is
accomplished by two factorizations:
\be \label{eq:twofact}
\begin{aligned}
\frac14(4\lambda_2^3+27\lambda_3^2) &= b(\lambda)d(\lambda), \\
\frac14(4\mu_2^3+27\mu_3^2) &= b(\mu) d(\mu).
\end{aligned}
\ee
In other words (considering the vanishing loci), the discriminant
locus $\Delta_\tau$ of the first fibration is decomposed into two
divisors $\Delta'_\tau = \{b(\lambda)=0\}$ and
$\Delta''_\tau = \{d(\lambda)=0\}$, and similarly for $\Delta_\rho$.
It follows that $b(\lambda)$, $d(\lambda)$, $b(\mu)$, $d(\mu)$ are
sections of
\be
\mathcal{O}(\Delta'_\tau), \mathcal{O}(\Delta''_\tau),
\mathcal{O}(\Delta'_\rho),
\mathcal{O}(\Delta''_\rho),
\ee
respectively.
We can write the equation for the F-theory dual in the
form\footnote{Here we are using the fact that
\be
Y^2 = X^3 + a \sigma^4 X + b\sigma^5 + c \sigma^6 + d\sigma^7
\ee
is birational to
\be
y^2 = x^3 + (s^3 + as + c)x^2 + bd x.
\ee
}
\be Y^2 = X^3   -3\lambda_2\mu_2 \sigma^4 X + b(\lambda) b(\mu)\sigma^5
- \frac{27}2\lambda_3\mu_3 \sigma^6 + d(\lambda) d(\mu)\sigma^7.
\ee

Again, we can determine line bundles from coefficients:
\be
\begin{aligned}
\mathcal{O}(4L_\tau + 4 L_\rho)
  &= \mathcal{O}(-4K_B),\\
\mathcal{O}(\Delta'_\tau+\Delta'_\rho)
  &= \mathcal{O}(-6K_B+\Lambda_{(E_8\times E_8)\rtimes\mathbb{Z}_2}),\\
\mathcal{O}(6L_\tau + 6L_\rho)
  &= \mathcal{O}(-6K_B),\\
\mathcal{O}(\Delta''_\tau+\Delta''_\rho)
  &= \mathcal{O}(-6K_B-\Lambda_{(E_8\times E_8)\rtimes\mathbb{Z}_2}).
\end{aligned}
\ee
Note that
\be
\mathcal{O}(\Delta''_\tau+\Delta''_\rho)=
\mathcal{O}(12L_\tau-\Delta'_\tau+12L_\rho-\Delta'_\rho)=
\mathcal{O}(-12K_B-\Delta'_\tau-\Delta'_\rho),
\ee
so the second and fourth equations above are equivalent.

It follows that (up to torsion):
\be
\begin{aligned}
\mathcal{O}(L_\tau + L_\rho)
  &= \mathcal{O}(-K_B),\\
\mathcal{O}(\Delta'_\tau + \Delta'_\rho)
  &= \mathcal{O}(-6K_B+\Lambda_{(E_8\times E_8)\rtimes\mathbb{Z}_2}).
\end{aligned}
\ee
Thus, for a given base $B$,
we will be able to construct a
non-geometric compactification for the $(E_8\times E_8)\rtimes\mathbb{Z}_2$
heterotic string out of any two elliptic fibrations $\pi_\tau:\mathcal E\to B$
and $\pi_\rho:\mathcal F\to B$, together with decompositions of their
discriminant divisors
\be
\Delta_\tau = \Delta'_\tau + \Delta''_\tau
\quad \text{and} \quad
\Delta_\rho = \Delta'_\rho + \Delta''_\rho,
\ee
provided that the associated line bundles
$\mathcal{O}(L_\tau)$ and $\mathcal{O}(L_\rho)$ satisfy
\be \label{eq:keyE8}
\mathcal{O}(L_\tau+L_\rho)=\mathcal{O}(-K_B),
\ee
up to torsion.

\subsection{Compactifications to six dimensions}

In six dimensions, it is possible to choose $B=T^2$ with
$\mathcal{O}(\Lambda_{\operatorname{Spin}{32}/\mathbb{Z}_2})$,
$\mathcal{O}(L_\tau)$ and $\mathcal{O}(L_\rho)$
all being torsion line bundles.  This leads to the
familiar compactification
on $T^2\times T^2$, or orbifolds thereof, and is not a case we will analyze
in detail.  In particular, both $\tau$ and $\rho$ are constant in this case,
and the heterotic model is geometric.

The other possibility in six dimensions is $B=\mathbb{P}^1$, and there
are then three cases (bearing in mind that the Picard group has no
torsion in this case), stemming from the formula
$\mathcal{O}(L_\tau+L_\rho)=\mathcal{O}(-K_B)$, together with the
fact that $\mathcal{O}(4L_\tau)$, $\mathcal{O}(6L_\tau)$, $\mathcal{O}(4L_\rho)$, and $\mathcal{O}(6L_\rho)$ all have sections:
\begin{enumerate}
\item $\mathcal{O}(L_\tau)$ has degree $2$
which implies that $\mathcal{O}(L_\rho)$
is trivial and hence that $\rho$ is constant.  This is a geometric model
in which $\mathcal E$ is an elliptically fibered $K3$ surface.
\item $\mathcal{O}(L_\tau)$ and $\mathcal{O}(L_\rho)$
each have degree one.  This implies that
both $\mathcal E$ and $\mathcal F$ are rational elliptic
surfaces.\footnote{These
are sometimes called ``$dP_9$ surfaces.''}
Both $\tau$ and $\rho$
are non-constant; these are the Hellerman--McGreevy--Williams models.
\item $\mathcal{O}(L_\tau)$ is trivial, and $\mathcal{O}(L_\rho)$
has degree $2$.  In this case,
$\tau$ is constant but $\rho$ varies; this is (fiberwise) mirror symmetric to
case (1).
\end{enumerate}

In case (1), we recover the familiar geometric compactifications and their
known F-theory duals.  There is one additional feature of these models
which we now spell out in detail: in order for the heterotic gauge group
to remain unbroken, all instantons must be point-like, and as such, each
must be located at a particular point on the heterotic side.  As our basic
construction shows, the complex structure of the F-theory model is determined
by the $\rho$ and $\tau$ data on the heterotic side, and appears to be
independent of
the location of the small instantons.  However, each complex structure modulus
on the F-theory side is part of a hypermultiplet which includes an additional
complex scalar, and it is those scalars which dictate the locations
of the small instantons.  In a typical vacuum, the expectation values of
those scalars vanish, so one would expect there to be a preferred location
for small instantons.

In the $\operatorname{Spin}{32}/\mathbb Z_2$ case, the physics of small
instantons was described by Witten \cite{Witten:1995gx};
Aspinwall \cite{Aspinwall:1996vc} used this analysis to identify
the corresponding features of F-theory:
small $\operatorname{Spin}{32}/\mathbb Z_2$ instantons
correspond to zeros of the coefficient $\varepsilon$ in the basic equation
\eqref{eq:AMbis}.  Aspinwall also gave an explanation of the
zeros of $p_0$ in \eqref{eq:AMbis}: they correspond to ``hidden obstructors''
\cite{sixauthors} which occur at singular points of the heterotic K3 surface.
As already mentioned, we do not consider hidden obstructors in our analysis
and in fact we have set $p_0=1$.

Since we have
\be
\varepsilon=\frac1{16}(4\lambda_2^3+27\lambda_3^2)(4\mu_2^3+27\mu_3^2),
\ee
and since $4\mu_2^3+27\mu_3^2$ is constant in case (1), we see that
the zeros of $\varepsilon$ correspond to the singular fibers of the
elliptic fibration $\mathcal E\to B$ (whose total space is the heterotic
K3 surface).  It is
natural to suppose that the
small instantons must be located along those singular fibers; in
fact, the most natural place to locate these small instantons is at
the singular points of the singular fibers.
Similar remarks apply to case (3), using $\mathcal F\to B$ instead of
$\mathcal E\to B$.

For the non-geometric compactifications, the zeros of $\varepsilon$
correspond to places where {\em either}\/ the fiber of $\mathcal E\to B$
is singular, {\em or}\/ the fiber of $\mathcal F\to B$ is singular.
The geometric part of this compactification is captured by $\mathcal E\to B$,
whose total space is the heterotic rational elliptic surface: the bundle
on this surface should have $12$ small instantons, so we can again locate
them at the singular points of fibers of $\mathcal E\to B$.  The additional
zeros of $\varepsilon$ correspond to singular fibers of $\mathcal F\to B$
and don't have a straightforward geometric interpretation.

A similar analysis applies when the gauge group is
$(E_8\times E_8)\rtimes \mathbb Z_2$.
This time, the small instantons involve tensionless strings
\cite{GanorHanany,SeibergWittensix}, and in case (1) we
must choose how to distribute $24$ small instantons between the two
$E_8$ factors.  In
the F-theory interpretation \cite{FCY1}, the zeros of the coefficients
$b$ and $d$ in the basic equation \eqref{eq:MV} correspond to the small
instantons in the two different $E_8$ factors.  In our construction,
the factorizations \eqref{eq:twofact} show that the zeros of $b$ and $d$,
together, correspond to the singular fibers of the two elliptic fibrations
$\mathcal E\to B$ and $\mathcal F\to B$.  Thus, in case (1) the $24$ singular
fibers of $\mathcal E\to B$ get divided into two groups, corresponding to
the two $E_8$ factors.  As in the $\operatorname{Spin}{32}/\mathbb Z_2$
case, we propose that the small instantons should be located at the singular
points of those singular fibers.  Case (3) is similar, with the
roles of $\mathcal E\to B$ and $\mathcal F\to B$ reversed.

In case (2), the singular fibers of $\mathcal E\to B$ get divided
into two groups, according to \eqref{eq:twofact}, and the singular
fibers of $\mathcal F\to B$ likewise get divided.  The total space
of $\mathcal E\to B$ is the rational elliptic surface upon which we
are compactifying the heterotic string, and we locate the $12$ small
instantons at the singular fibers of $\mathcal E\to B$, divided into
two groups as in \eqref{eq:twofact}.  The additional zeros,
corresponding to singular fibers of $\mathcal F\to B$, again do not
have a straightforward geometric interpretation.

One interesting thing to note is that no new F-theory models were
required in six dimensions to provide duals for non-geometric
compactifications:  all of the duals to non-geometric
compactifications are in the same class of F-theory models as the
duals to {\em geometric}\/ compactifications, although presumably
the dualities are occurring at different locations in the
hypermultiplet moduli space. In four dimensions, some of
the semi-classical moduli are lifted by
fluxes~\cite{Dasgupta:1999ss}, so there may indeed be different
F-theory models for geometric and non-geometric compactifications in
that dimension.

\subsection{Compactifications to four dimensions: an example}

The general procedure described in section~\ref{constrnongeom}
can also be used to construct examples in dimension four,
which on the F-theory side will involve $K3$-fibered Calabi--Yau
$4$-folds.  These are much less constrained than was the
corresponding set of $K3$-fibered Calabi--Yau $3$-folds which
we used in the previous section, so rather than attempting a
general classification we will settle for examples.
Our examples are easily generalizable to
arbitrary $K3$-fibered Calabi-Yau $4$-fold and our results are characteristic of the general construction.

The class of examples we are interested in are Calabi-Yau $4$-folds, $\cM_4$, with a $\PP^2$ base which admit a $K3$-fibration. Schematically:
\begin{center}
\begin{tabular}{lcr}
\xymatrix{
 K3  \ar[r] & \cM_4 \ar[d] \\
&  \wt\PP^2 \\
}
&~~~~~with~~~~~&
\xymatrix{ T^2  \ar[r] & K3 \ar[d] \\
 & \PP^1 \\
}
\end{tabular}
\end{center}
The elliptic curve is represented as a hypersurface in $\PP^2$ via the vanishing of the Weierstrass equation \C{eq:weierstrass}. Let $[x,y,z]$ denote the homogeneous coordinates for this $\PP^2$, $[t_1,t_2,t_3]$ the homogeneous coordinates for the base $B=\wt\PP^2$, $[s_1,s_2]$ the coordinates for the $\PP^1$.
We construct a variety $S$ which is fibered over $B=\wt\PP^2$ with fiber $\PP^1$,
and a variety
$\mathbb{P}(\mathcal{O}\oplus\mathcal{O}(2L)\oplus\mathcal{O}(3L))$
which is fibered over $S$ with fiber $\PP^2$.
The varieties $\PP^1$ and $\PP^2$ are fibered over the base $\wt \PP^2$. We do this by lifting the torus action used to construct the base to act on the fiber: the coordinates $[x,y,z]$ and $[s_1,s_2]$ become sections of certain line bundles. Using $\lambda$ and $\mu$ to denote the $\CC^*$ actions of the base $\wt \PP^2$ and $\PP^1$ respectively we consider
\bea
[t_1,t_2,t_3] &\sim& \lambda [t_1,t_2,t_3],\cr
 [s_1,s_2] &\sim& [\lambda^{A_1}\mu s_1,\lambda^{A_2}\mu s_2], \cr
  [x,y,z] &\sim& [\lambda^{B_1} \mu^{C_1} x, \lambda^{B_2}\mu^{C_2} y, z],\label{eqn:torus_action_1}
\eea
with $A_i,B_i,C_i$ some real positive constants to be determined. Note we have made a basis choice such that the torus action on $z$ is trivial.
We require that \C{eqn:torus_action_1} acts consistently on the Weierstrass polynomial (written here in the homogeneous coordinates of $\PP^2$):
\be
P = -y^2 z + x^3 + z^2 x f(s,t) + z^3 g(s,t),\label{eqn:w_1}
\ee
and require that the variety defined by $P=0$ have trivial canonical class;
these conditions determine $B_i,C_i$ in terms of $A_i$. Picking $A_1=n,A_2=0$, it is convenient to write the exponents of the three $\CC^*$ torus actions defining $\wt\PP^2, \PP^1, \PP^2$ as a matrix:
 \be
 \begin{pmatrix}
 t_1 & t_2 & t_3 & s_1 & s_2 & x & y & z\\
 1 & 1 & 1 & n & 0 & 2(3+n) & 3(3+n)& 0\\
 0 & 0 & 0 & 1   &   1 & 4 & 6 & 0\\
 0 & 0 & 0 & 0   &   0 & 1 & 1 & 1\\
 \end{pmatrix}.\label{eqn:charge_1}
 \ee
This is precisely the charge matrix of a linear sigma model describing a toric variety~\cite{Witten:1993yc}.  Also $f(s,t)$ has charge $4(3+n)$ and $8$ under the first two $\CC^*$ actions, while $g(s,t)$ has charge $6(3+n)$ and $12$. This gives a class of $4$-folds with a twist labeled by the integer $n$.
We focus on F-theory duals of compactifications of the $(E_8\times E_8)\rtimes\mathbb Z_2$
heterotic string with unbroken
$(E_8\times E_8)\rtimes\mathbb Z_2$; then the twist parameter
$n$ corresponds to choosing the line bundle ${\mathcal O}(\Lambda_{(E_8\times E_8)\rtimes\mathbb Z_2})$
to be ${\mathcal O}_{\wt\PP^2}(n)$.
In order to get unbroken $(E_8\times E_8)\rtimes \mathbb Z_2$, we need to restrict to $n\le3$.

Unbroken $(E_8\times E_8)\rtimes\mathbb Z_2$ implies that the
Weierstrass equation takes the special form \eqref{eq:MV} with
the coefficients interpreted as sections of appropriate bundles.
In our case, this implies that
\[ f(s,t) = a(t) s_1^4s_2^4 \]
and
\[ g(s,t) = b(t) s_1^5s_2^7 + c(t) s_1^6 s_2^6 + d(t) s_1^7s_2^5\]
where $a(t)$, $b(t)$, $c(t)$ and $d(t)$ are homogeneous of degrees
$12$, $18+n$, $18$, and $18-n$, respectively.\footnote{Note that any
model in dimension four with unbroken $(E_8\times E_8)\rtimes\mathbb Z_2$
has confusing aspects,
such as an infinite tower of light solitonic states,
 if $b(t)=0$ intersects $d(t)=0$ \cite{codimthree};
such an intersection is unavoidable for our choice $B=\mathbb P^2$.}

To find F-theory duals for geometric or non-geometric heterotic
compactifications, following eq.~\eqref{eq:keyE8} we must choose line
bundles $\mathcal{O}(L_\tau)$ and $\mathcal{O}(L_\rho)$ of degrees
$d$ and $3-d$ (since $\mathcal{O}(-K_B)$ has degree $3$), as well
as a decomposition of the divisor $\Delta_\tau$ (of degree $12d$) into
two components $\Delta_\tau' + \Delta_\tau''$ of degrees $k$ and
$12d-k$, and
 a decomposition of the divisor $\Delta_\rho$ (of degree $36-12d$) into
two components $\Delta_\rho' + \Delta_\rho''$ of degrees $\ell$ and
$36-12d-\ell$, such that $k+\ell = 18+n$.

The choice of $\mathcal{O}(L_\tau)$ and $\mathcal{O}(L_\rho)$ presents
no particular problem for any value of $d\in\{0,1,2,3\}$, so there
are a variety of geometric and non-geometric heterotic compactifications
with F-theory duals of this kind.  In fact,
there are equal numbers of geometric
and non-geometric models (treating the constant $\tau$ models as geometric,
even though strictly speaking they are the mirrors of geometric models),
so neither type of heterotic compactification is favored.  That is,
not only are non-geometric models possible, they are just as typical as
geometric models.

Note that the choice of decompositions
$\Delta_\tau = \Delta_\tau' + \Delta_\tau''$
and $\Delta_\rho = \Delta_\rho' + \Delta_\rho''$,
which affects the distribution of instantons (and their non-geometric
counterparts) between the two  $E_8$ factors of the gauge group,
 is trickier: the Weierstrass
equations describing the bundles $\mathcal E$ and $\mathcal F$
must be carefully tuned to guarantee such a decomposition.
This does not, however, affect our discussion of typicality for
geometric versus non-geometric heterotic compactifications.

The description of moduli spaces which follows from the above
analysis is a semi-classical one, and in general we expect a number
of moduli to be lifted by fluxes~\cite{Dasgupta:1999ss}.  As a
consequence, we should expect that the F-theory duals of geometric
and non-geometric compactifications live on different (quantum)
moduli spaces in dimension four.  It would be interesting to have a
concrete example of this phenomenon.

\subsection{Tadpoles and the Bianchi identity}
\label{tadpoles}

One of the beautiful features of this class of heterotic models is
that there is no need for extra ingredients like orientifold planes
to construct compact models. Instead, the Bianchi identity for
$\mathcal{H}_3$-flux, \be  d {\cal H}_3 = { \frac{\alpha'}4} \left(
\Tr (R \wedge R) - \Tr (F \wedge F) \right), \label{tad} \ee
 includes a higher derivative correction  that in the presence of
 curvature induces a five-brane charge thereby allowing one
 to construct compact solutions. In the geometric setting, the five-brane charge tadpole must
 be satisfied by a combination of wrapped NS5-branes and finite-size
 gauge instantons.

 We would like to understand this tadpole in the more
 general non-geometric setting. This is a subtle question for
 reasons that we will outline and is really best
 answered from a world-sheet approach.

 Let us first think adiabatically from the perspective of a
 physicist who has reduced on the torus fiber and observes physics purely on
 the base $B$. From the perspective of such an observer, there are
 two scalar fields $\tau$ and $\rho$ with monodromies around
 divisors of $B$. If Wilson line moduli were included, they would
 give additional scalars with the entire collection acted on by
 the full heterotic $T^2$ duality group. For simplicity, we will
 continue to restrict to unbroken maximal gauge symmetry.

First, our usual intuition is that the total $\rho$ monodromy measured
around any divisor in $B$ must be trivial for a compact solution.
Said differently: the total NS5-brane charge must vanish. It is
worth pointing out that this is not required in the heterotic string
since the deficit can be made up by the gravitational contribution
to the charge.

Second, there are really several distinct cases to this adiabatic tadpole analysis depending on whether ${\cal H}_3$ has support along the torus fiber.
 If ${\cal H}_3$ has one or two legs along the fiber then the left hand side of~\C{tad}\ becomes
 intrinsically non-geometric, and would be sourced by some non-geometric analogue of a
 Pontryagin class. That is, the standard expression $\Tr (R \wedge R)$ is not
 invariant under the $SL(2,\Z)$ action on $\rho$ and so is not
 well-defined. To understand these components of the tadpole
 requires a world-sheet analysis. While this case does not seem to occur for models with F-theory duals,
it would be very interesting to determine whether a non-trivial
$d{\cal H}_3$ could be sourced this way since it would provide a new
kind of non-K\"ahler solution which locally satisfies the quite
restrictive supersymmetry constraints, while solving the Bianchi
identity~\C{tad}\ via T-duality.

The only component of $d{\cal H}_3$ for which we might be able to
use our adiabatic picture is when $d{\cal H}_3$ is supported
completely on $B$, which requires a compactification to four
dimensions or lower. This is the charge for NS5-branes which wrap
the torus fiber. We note that $\rho$ monodromies can never create
NS5-branes which wrap the fiber. Those branes are always transverse
to the torus fiber. So at least for this component, we might hope to
treat $\tau$ and $\rho$ in a similar fashion.

 Let us try a direct attempt to understand this component of the charge tadpole. As a
 warm up case, let us take a geometric heterotic compactification
 on an elliptic space $\M \rightarrow B$. What we would like to do
 is express the Chern classes of $\M$ in terms of those of $B$
 together with the data defining the elliptic fibration. We can
 follow an approach used in~\cite{SVW,Friedman:1997yq,  Robbins:2004hx}. For this geometric model
 we can present $\M$ in
 Weierstrass form as before. Let $W$ be a $\PP^2$ bundle over $B$
 with homogeneous coordinates $[u,v,w]$ which are sections of
 \be\mathcal{O}(1) \otimes \mathcal{O}(2 L_\tau), \quad \mathcal{O}(1)\otimes \mathcal{O}(
 3 L_\tau), \quad \mathcal{O}(1), \ee respectively. The line bundle $\mathcal{O}(1)$ is the degree one
 bundle over the $\PP^2$ fiber. To describe $\M$, we consider
 \be
s = -wv^2 + u^3+ \lambda_2 u w^2 + \lambda_3 w^3=0,
 \ee
where $\lambda_2$ and $\lambda_3$ are sections of the line bundles
$\mathcal{O}(4 L_\tau)$ and $\mathcal{O}(6 L_\tau)$, respectively
while $s$ is a section of $\mathcal{O}(3) \otimes \mathcal{O}(6
L_\tau)$. In this purely geometric model, we set \be
\mathcal{O}(L_\tau) = \mathcal{O}(-K_B), \ee to ensure that $\M$ is
Calabi-Yau.

Let us set $\alpha = c_1(\mathcal{O}(1))$. The cohomology ring of
$W$ is then generated over the cohomology ring of $B$ by the
addition of $\alpha$ together with the relation, \be \alpha
(\alpha+2c_1(B))(\alpha+3c_1(B))=0. \ee This relation
states that $(u,v,w)$ are not permitted to have any common zeroes.
This relation holds in the cohomology ring of $W$. To restrict to
$\M$, we want to impose $s=0$ but $s$ is itself a section of a
bundle with first Chern class $3(\alpha+2c_1(B))$. Any class on $\M$
that can be extended to $W$ can be integrated over $\M$ by
multiplying by $3(\alpha+2c_1(B))$.

Now we are ready to compute the Chern classes of $\M$ in terms of
those of $B$. Let $C_B$ denote the total Chern class of $B$. The
total Chern class $C_W$ of $W$ is \be C_W = C_B \cdot (1+\alpha)(1+
\alpha+2c_1(B))(1+\alpha+3c_1(B)). \ee To get the Chern class of $\M$, we
use adjunction: \be \label{CM} C_\M = C_W \cdot \frac{1}{1+
3(\alpha+2c_1(B))}.\ee To compute the five-brane tadpole~\C{tad}, we
are really interested in $p_1(\M)$ so we want to extract $c_1(\M)$
and $c_2(\M)$ from~\C{CM}. It is easy to check that in this case,
$c_1(\M)=0$ as we expect. On expanding, we find \be
\label{geomanomaly} c_2(\M) = c_2(B) + 4 \alpha c_1(B) + 11 c_1(B)^2. \ee

How might this computation generalize to include $\rho$ monodromies?
It is important to note that the choice of connection used to
compute $\Tr (R \wedge R)$ is quite central. The connection required
by duality is the torsional connection \be \Omega_+ = \omega +
\frac12 {\cal H}_3, \ee where $\omega$ is the spin connection;
see~\cite{Becker:2009df, Becker:2009zx}\ for a discussion about the
role of the connection in constructing geometric torsional
solutions.

At the level of cohomology classes, the choice of connection does
not matter -- at least for geometric backgrounds. For non-geometric
backgrounds, the torsional connection will certainly depend on
$\rho$, and this dependence might now involve non-trivial topology
induced from $\rho$ monodromies. In a patch where $\rho$ and $\tau$
are single-valued, the metric itself depends on $\rho_2$ via the
combination $\rho_2/\tau_2$ in~\C{fiberedmetric}\ while a dependence
on $\rho_1$ emerges from ${\cal H}_3$ in the connection. From these arguments, it seems clear that $\rho$ will contribute to
the gravitational source for the tadpole although the precise form
of the contribution is unknown. To proceed, let us treat $\tau$ and
$\rho$ symmetrically as an ansatz. This is somewhat suggested both
by duality with F-theory and by a mirror transform on the fiber
which exchanges $\rho$ and $\tau$ but leaves wrapped NS5-branes
invariant.

So rather than a single $\PP^2$ bundle over $B$, let us consider a
$\PP^2 \times \PP^2$ bundle over $B$. We will take a cubic surface
in each $\PP^2$ with one encoding the $\tau$ variation and the other
encoding the $\rho$ variation, very much in the spirit of the
doubled torus formalism. Let $\beta = c_1(\mathcal{O}(1))$ for the
second $\PP^2$. We will impose our earlier constraint~\C{eq:keyE8}\
that $$ \mathcal{O}(L_\tau+L_\rho)=\mathcal{O}(-K_B)$$ and the
relations \bea \label{ringrel} && \alpha(\alpha + 2
c_1(L_\tau))(\alpha + 3 c_1(L_\tau))=0, \non
\\ && \beta(\beta + 2 c_1(L_\rho))(\beta + 3
c_1(L_\rho))=0, \eea where we have abbreviated $c_1(\mathcal{O}(L))$
by $c_1(L)$ to reduce notational clutter. Because there is really
only one physical torus fiber, it only really makes sense to
integrate out the fibers and discuss the anomaly on the base. Given
this aim, we can simplify the relations~\C{ringrel}\ to \bea
\label{finalring} && \alpha(\alpha + 3 c_1(L_\tau))=0, \non \\ &&
\beta(\beta + 3 c_1(L_\rho)) =0. \eea Now the analogue of~\C{CM}\
becomes \bea \label{CMnongeom} C_\M &=& C_B \cdot
 (1+\alpha)(1+ \alpha+2c_1(L_\tau))(1+\alpha+3c_1(L_\tau)) \times \non\\   &&
\frac{(1+\beta)
 (1+ \beta+2c_1(L_\rho))(1+\beta+3c_1(L_\rho))}{
 [1+ 3(\alpha+2c_1(L_\tau))][1+ 3(\beta+2c_1(L_\rho))]}.\eea
First it is easy to check that $c_1(\M)=0$ simply because
of~\C{eq:keyE8}\ and the linearity of the computation of $c_1$. This
is completely natural. The more interesting structure is $c_2$ which
is non-linear. We now find \bea \label{nonchern} c_2(\M) &=& c_2(B)
+ 11 c_1(B)^2 - 95 c_1(L_\tau) c_1(L_\rho) - 9 \alpha\beta \non\\ &&
- 36(\beta c_1(L_\tau) + \alpha c_1(L_\rho)) + 4(\alpha c_1(L_\tau)
+ \beta c_1(L_\rho)). \eea The first three terms of~\C{nonchern}\
can be directly compared with~\C{geomanomaly}\ since they are fully
supported on the base. The interesting addition is the quadratic
term $- 95 c_1(L_\tau) c_1(L_\rho)$ which is only present in the
non-geometric case. While this computation suggests this coupling is
present, it would be very interesting to understand whether this is
true directly from a world-sheet computation.

It would also be nice to compare the NS5-brane anomaly to the
D3-brane anomaly of F-theory~\cite{SVW}, as was done
in~\cite{Friedman:1997yq}\ for a class of dual pairs. The quantity
to be determined in F-theory, namely the D3-brane charge, is
unambiguous though the singularities of the F-theory four-fold,
reflecting the unbroken maximal gauge symmetry, make that
computation potentially subtle. What is far less clear is what that
number should be compared with in the heterotic string. In the
geometric setting, U-duality related NS5-branes wrapping the
elliptic fiber to D3-branes rather directly but that chain is
certainly modified by the presence of $\rho$ monodromies.


For models admitting an F-theory dual, an M5-brane wrapped on the
$K3$-fiber of the F-theory geometry does naturally provide a
realization of the world-sheet of the non-geometric heterotic
string. This is similar to the proposal in \cite{Sethi:2007bw}\ for
studying sigma-models of (geometric) heterotic torsional
backgrounds.   We will not explore these interesting directions
here, leaving them to future work.

\section{Heterotic solutions with torsion}
\label{hettorsion} In the previous sections, we constructed
non-geometric heterotic solutions by solving the Bianchi identity
with point-like instantons -- the only way the $\mathcal{H}_3$ flux
appeared was via $\rho$ monodromies. Yet the most physically
interesting heterotic backgrounds involve more general torsion, or
$\mathcal{H}_3$ flux, since they contain fewer moduli than
conventional Calabi-Yau compactifications. From an F-theory
perspective, the simplification we used in the preceding discussion
is equivalent to setting any bulk filling $G_4$-flux to zero and
looking for heterotic duals of the F-theory geometry. In this
section, we wish to analyze the role of the bulk filling $G_4$-flux,
and its various dual descriptions. Some notation and general
relations of use in the following sections can be found in
Appendix~\ref{appendixnotation}.

Torsional solutions were first described by
\cite{strominger-torsion}\ in the context of supergravity, and  the
known compact examples are based on the solutions constructed in
\cite{Dasgupta:1999ss}. These geometries  are constructed by
dualizing M-theory compactified on
four-folds\footnote{In contrast to the prior discussion,
 the four-folds discussed in the next two sections
 are not necessarily Calabi-Yau. Hence, we will denote them by $\cM_4$ and not $\overline{Z}$.}, $\cM_4$, with
bulk filling $G_4$, resulting in a four-dimensional heterotic
compactification on a complex but non-K\"ahler geometric space with
non-trivial $\H_3$. The solution of the Bianchi
identity~\C{bianchi}\ is guaranteed by satisfying the tadpole
condition in M-theory \be n_{M_2} =  - \int_{\cM_4} X_8 - \half
\int_{\cM_4} G_4 \w G_4, \label{eqn:tadpole_1} \ee where
$n_{M2}$ is the number of space-time filling M2-branes. The integral
of $X_8$ is given by the Euler character of the four-fold
$\cM_4$:
$$  - \int_{\cM_4} X_8  = \chi/24.$$
In the language of section~3, those models involved $\tau$
monodromies but constant $\rho$. Using a duality chain shown in
figure~\ref{fig:duality_1}, we will show the presence of torsional
flux gives us an additional way to generate non-geometric solutions.
Dualizing flux to get ``non-geometric fluxes'' has been explored in
past work like~\cite{Shelton:2005cf, Shelton:2006fd}

These torsional non-geometric heterotic solutions, in turn, have
novel type IIB and M-theory duals, which we construct in section~5.
This basically completes a duality chain which starts with M-theory
on a conformally Calabi-Yau four-fold and generates new compact
solutions via U-duality.

We will not work in generality; rather we will focus on a simple
example that will illustrate most of the germane features. The main
simplification we use is an orbifold metric for a $K3$ surface. The
advantage of this replacement is that the the orbifolded theory
inherits part of the U-duality group of the covering toroidal
compactification. Otherwise, we would need to worry about patching
with mirror transforms of a $K3$ surface rather than the U-duality
group of a torus.
\begin{figure}[htp]
\centering \xymatrix{ {\rm M\!\!-\!\!theory}~{\rm on}~\cM_4 ~~~~
\ar[r] & ~~~{\rm type~I}~~~ \ar[r]^-{\rm S-duality}& ~~~{\rm
Heterotic}
\ar[r]^-{\rm T-duality}~~~ & ~~~{\rm Heterotic}~~~ \\
} \caption{Schematic of the duality chain that we use to generate
non-geometric heterotic solutions with flux.}\label{fig:duality_1}
\end{figure}

\subsubsection*{M-theory on a Calabi-Yau four-fold $\cM_4$}
Let us briefly review the main  differences between the M-theory
compactifications discussed in section~3 and compactifications with
bulk filling $G_4$-flux. As described in~\cite{Becker:1996gj}, with
flux the metric becomes warped so that the four-fold is now
conformally Calabi-Yau, \be ds^2 = e^{-\phi}\eta_{\mu\nu}dx^\mu
dx^\nu + e^{\half\phi} \wt g_{MN} dx^M dx^N,\label{eqn:metric_1} \ee
where $\wt g_{MN}$ is the metric on the four-fold.  The flux must be
a primitive $(2,2)$-form: \be G_{abcd}=G_{abc\bar d}=0, ~~ g^{c\bar
d} G_{a\bar b c \bar d} = 0. \ee There is also a space-time filling
flux: \be G_{\mu\nu\rho a} = \del_a e^{-3\phi/2}. \ee The warp
factor obeys a Poisson equation \be \Box e^{3\phi/2} =
\star_8\left\{4\pi^2 X_8 - \half G_4 \w G_4\right\} -  4\pi^2
\sum_{i=1}^n \delta^8(x-x_i),\label{eqn:laplacian_1} \ee where the
Laplacian and Hodge dual are taken with respect the unwarped
internal metric $g_{MN}$,  and we have allowed for the possibility
of space-time filling M2-branes localized at points, $x_i$, in the
four-fold.

There is an obstruction to solving the Poisson
equation~\C{eqn:laplacian_1}\ unless the charge cancelation
condition~\C{eqn:tadpole_1}\ is satisfied; the presence of the
higher derivative coupling, $X_8$, is crucial for the existence of a
solution. In our subsequent discussion, we will frequently arrive at
expressions that depend on higher derivative terms which, although
vanishing at the level of supergravity, are essential for the
existence of the solution.

These M-theory solutions have three-dimensional heterotic duals when
$\cM_4$ admits a $K3$ fibration, and four-dimensional duals when
that $K3$ fibration admits a compatible elliptic fibration. The
duality chain sketched in figure~\ref{fig:duality_1} involves
T-duality so we need a starting four-fold metric with suitable
(approximate) isometries.

We could start with a semi-flat metric~\C{fiberedmetric}\ for an
elliptic $4$-fold. However, to make our life simpler and still
illustrate the pertinent features of our solutions, we will further
restrict to a particularly nice four-fold; namely, $K3\times
\wt{K3}$, where both $K3$ spaces admit elliptic fibrations. Further,
we will take $\wt{K3}$ to be an orbifold space $\wt{K3} = T^4/\Z_2$.
On shrinking the elliptic fiber of $\wt{K3}$, we will arrive at the
orientifold limit of an F-theory compactification on $K3 \times 
\frac{T^2}{(-1)^{F_L}\, \Omega \, \Z_2 }$.

Let us take a square complex structure on $T^4$, and choose the
complex coordinates $(w,v)$ to have canonical periodicity. The
orbifold $\Z_2$ acts by sending \be (w,v)\rightarrow (-w,-v).\ee We
can choose $v$ to coordinatize the elliptic fiber and $w$ the base.
The four-form flux takes the form: \bea G_4 &=& \alpha \w dw \w
d\bar v + \beta \w d\bar w \w d\bar v + \cc, \eea where $\alpha\in
H^{1,1}(K3,\Z)$, $\beta\in H^{2,0}(K3, \Z)$ are primitive classes
with respect to the K\"ahler form of $K3$. If $\beta = 0$ then this
compactification preserves eight supersymmetries, otherwise it
preserves four supersymmetries.

\subsection{Type IIB and Type I torsional solutions}
\label{sect:typeI} As a first step to constructing the new heterotic
solutions, we take the F-theory limit by shrinking the elliptic
fiber with coordinate $v$. This gives a type IIB compactification on
$K3 \times \frac{T^2}{(-1)^{F_L}\, \Omega \, \Z_2 }$ with D7-branes
and possibly D3-branes, depending on whether $n_{M_2}$ is non-zero.
The metric for this background is given by, \be ds^2 = e^{-3\phi/4}
\eta_{\mu\nu}dx^\mu dx^\nu + e^{3\phi/4} g_{mn} dx^m dx^n +
e^{3\phi/4} |dw_1 + i dw_2|^2,\label{eqn:typeIIB_metric_1} \ee where
$dw = dw_1 + i dw_2$ is along the $T^2$ while the indices
$m,n=1,\ldots,4$ parametrize the directions along the $K3$ surface,
and $g_{mn}$ is the Ricci-flat $K3$ metric. The  M-theory $4$-form
flux lifts to type IIB $3$-form fluxes given by \bea H_3 = (\alpha
+\overline{\beta})\w dw + \cc, \quad F_3 = i(\overline{\beta} -
\alpha) \w dw + \cc,\non \eea and a $5$-form flux that fills
space-time \be F_5 = dC_4 + H_3 \w C_2, ~~{\rm where}~~ dC_4=\ve_4
\w d e^{-3\phi/2}. \ee We will often find it convenient to write \be
F_3 = F_{w^1} dw^1 + F_{w^2} dw^2= F_{w} dw + F_{\bar w} d\bar w,
\ee and similarly for $H_3$. Writing $\alpha = \alpha_1 + i
\alpha_2$ and $\beta = \beta_1 + i \beta_2$ we find \be
\label{eqn:iibfluxes_1}
\begin{aligned}
 H_3 &= 2(\alpha_1+\beta_1)\w dw^1 + 2(\beta_2
- \alpha_2)\w dw^2,\cr F_3 &= 2(\alpha_2+\beta_2)\w dw^1 +
2(\alpha_1 - \beta_1)\w dw^2,
\end{aligned}
\ee
 The fluxes
satisfy a constraint (corresponding to imaginary self-duality of
$G_3$) given by \bea F_3  = \star_6 (e^{-\Phi_B} H_3),\label{isd}
\eea where $\star_6$ is with respect to the unwarped metric. The
type IIB dilaton $\Phi_B$ is determined by the complex structure of
the elliptic fiber with coordinate $v$. We set $g_s= e^{\Phi_B}$. If
$\beta = 0$ then $H_{w^1} = g_sF_{w^2}$, and $H_{w_2} = - g_s
F_{w_1}$.

By T-dualizing along the $w_1,w_2$ coordinates, we arrive at a
geometric type I configuration with flux. This type I solution
consists of a six-dimensional manifold that is torus fibered with
metric \be ds^2 = e^{-3\phi/4} \eta_{\mu\nu}dx^\mu dx^\nu +
e^{3\phi/4} g_{mn} dx^m dx^n + e^{-3\phi/4}|dw  +
A_H|^2.\label{eqn:typeI_metric_2} \ee The one-form $A_H = B_{w^1} +
i B_{w^2}$ is constructed out of a trivialization of the type IIB
field strength $H_3$. The trivialization is chosen such that the
$B_{w^i}$ are independent of the $T^2$ elliptic fiber in the $K3$
surface. This is a gauge choice which is convenient for the next
step in the duality chain.

The only non-zero RR flux is \bea F_3' &=&  F_{w^1} \w dw^2 -
F_{w^2} \w dw^1 + (F_{w^1} \w B_{w^2} - F_{w^2} \w B_{w^1}) +
\star_{K3} d e^{3\phi/2},\label{typeIf3} \eea where in the last line
we used $(F_5)_{w_1 w_2} = -\star_{K3} d e^{3\phi/2}$. Note that
$dF_3' = 0$ at the level of supergravity, and $F_1'=F_5'=0$
consistent with the type I field content. These are the solutions
of~\cite{Dasgupta:1999ss}; a similar chain starting with an elliptic
Calabi-Yau $3$-fold in the semi-flat approximation gives more
general metrics described in~\cite{Becker:2009df}. Fortunately, we
can extract the physics we wish to see starting from this clean
example.

\subsection{New heterotic solutions with torsion}
\label{newhet}

We follow the duality sequence illustrated in
figure~\ref{fig:duality_1}. Start with the type I solution in
section \ref{sect:typeI} and S-dualize to the heterotic string. Then
apply two T-dualities along the fiber of the $K3$ factor to generate
the new non-geometric heterotic solution. This is the extra
ingredient and the remaining unexplored duality direction in the
possible dual realizations of F-theory on $K3\times \wt{K3}$.

If we choose an orbifold metric $T^4/\Z_2$ for this $K3$ factor, we
can write down explicit expressions for the metric and fluxes.
Again, we could take a more general semi-flat metric but this should
suffice.

So let us take a $K3$ surface realized as a Kummer surface
$T^4/\Z_2$; further choose $T^4 = T^2 \times T^2$. Let $z_1 = x_1 +
i y_1$ be coordinates for the $T^2$ fiber of the $K3$ surface, and
let $z_2 = x_2 + i y_2$ be coordinates of the $T^2/\Z_2$ base. For
simplicity, we choose square tori with canonical periodicities so
that $dz_i = dx_i+i dy_i$ with $i=1,2$ is a basis of holomorphic one
forms.

We will construct an $N=2$ solution, whose existence post-duality is
more trust-worthy, by choosing $\beta=0$ and $\alpha$ to be the
following $(1,1)$-form: \be \alpha = A dz_1 \w d\bz_2.\ee The
constant $A$ is real. With this choice, the fluxes can be
trivialized as follows (in real coordinates): \bea B_{w^1} &=& 2A
(x_2 dx_1 + y_2 dy_1), \quad B_{w^2} = 2A(y_2 dx_1 - x_2 dy_1),\cr
C_{w^1} &=& 2A (x_2 dy_1 - y_2 dx_1), \quad C_{w^2} = 2A(x_2 dx_1 +
y_2 dy_1).\label{eqn:fluxes_1} \eea We pick this trivialization to
ensure that there are isometries along the $(x_1,y_1)$ directions.
We can T-dualize along these directions to give a new heterotic
solution (we denote the new field components by hats): \bea \wh ds^2
&=& \eta_{\mu\nu}dx^\mu dx^\nu + e^{3\phi/2} [ \varpi^2 (dx_1^2 +
dy_1^2) + (dx_2^2 + dy_2^2)] + (dw_1 + \varpi B_{w^1})^2 + \cr
&&\qquad +  (dw_2 + \varpi B_{w^2})^2 , \eea where $\varpi$ is given
by \be \varpi = (e^{3\phi/2} + 4|A|^2(x_2^2 +
y_2^2))^{-1}.\label{varpi2} \ee Note that at the level of
supergravity, the solution to the warp factor equation
\C{eqn:laplacian_1} with this choice of fluxes is given by \be
e^{3\phi/2} = 1 - 4A^2 (x_2^2 + y_2^2) + \O(\alpha')\ee implying
$\varpi = 1 + \O(\alpha'^2)$. The remaining terms arise from higher
derivative corrections to the warp factor equation
\C{eqn:laplacian_1}\ needed to ensure a solution exists. The ${\cal
B}$-field is given by \be {\cal B}_{w^1} = -2A \varpi (x_2 dx^1 +
y_2 dy^1), \quad {\cal B}_{w^2} = -2A \varpi (y_2 dx^1 - x_2 dy^1),
\ee and the heterotic dilaton is given by \be e^{\Phi_h} =
e^{3\phi/4}\varpi. \ee This solution is non-geometric in the
following sense. Locally, the solution has a well-defined
supergravity description and the above field content solves the
supersymmetry conditions and equations of motion (we show this
explicitly in the type IIB and M-theory duals below). On the other
hand, the background is only globally well-defined when we include
the $\SO(4,4,\Z)$ transformations of the $T^4$ fiber. This is to be
contrasted with the mechanism for generating non-geometries
described in section 3. Since $\wt{K3}$ is trivially fibered over
$K3$, that approach would give a geometric heterotic background.

It would be very interesting to explore compactifications in which
this kind of torsion $\mathcal{H}_3$-flux and the $\rho$ monodromies
of section~3 are combined.

\section{Type IIB and M-theory non-geometric duals} \label{iibm}
\subsection{G-structures and local geometry}
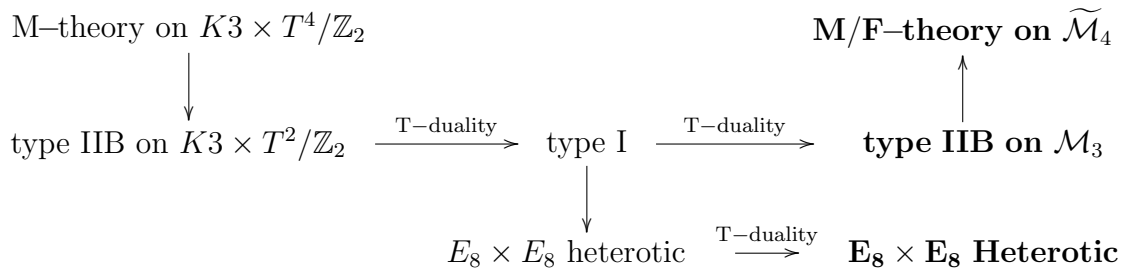
\begin{figure}[htp]
\begin{center}\leavevmode
\xymatrix{
{\rm M\!\!-\!\!theory}~{\rm on}~ K3 \times T^4/\Z_2 \ar[d]       &                 & {\rm \bf M/F\!\!-\!\!theory~on~}\wt\cM_4 \\
{\rm type~IIB~on~}K3 \times T^2/\Z_2 ~~\ar[r]^-{\rm T-duality} & ~~{\rm type~I}~~ \ar[d]  \ar[r]^-{\rm T-duality}& ~~~~ {\rm\bf type~IIB~on~}\cM_3 \ar[u] \\
 & E_8\times E_8~{\rm heterotic} ~~~~ \ar[r]^-{\rm T-duality} & ~~~~ {\bf E_8\times E_8~{\rm\bf Heterotic}}
} \caption{The duality chain used to generate the heterotic
solutions in the previous section as well as their type IIB and
M-theory dual descriptions. The bold face indicates new solutions
discussed in this paper. }\label{fig:duality_2}
\end{center}
\end{figure}

In this final section we will describe the dual type IIB  and
M-theory descriptions of the heterotic solutions derived in the
previous section (see figure \ref{fig:duality_2}). These solutions
are novel, having an interesting local and global geometry. We will
characterize the local geometry in terms of G-structures, developed
in \cite{Gauntlett:2002sc,Gauntlett:2002fz}, in which the spinors
classify the local geometry in terms of the fluxes; see, for
example,~\cite{Gauntlett:2003cy,Grana:2005jc}\ for reviews.

In compactifications without flux to four dimensions, supersymmetry
requires the existence of a covariantly constant spinor on the
internal six-dimensional manifold. This implies the holonomy group
is reduced to $\SU(3)$, which is a defining characteristic of a
Calabi-Yau manifold. The supersymmetry spinors can be used to form
spinor bilinears which correspond to forms on the internal space.
Two forms play a distinguished role:
 \be
 J_2 = -2i \eta^\dagger \gamma_{MN} \eta\, dx^M dx^N, \quad \Omega_3 = -2i \eta^T \gamma_{MNP} \eta dx^M dx^N dx^P.
  \ee
Since the spinor is covariantly constant, these forms are closed. It
is not hard to show that $J$ corresponds to the K\"ahler form and
$\Omega$ the holomorphic $3$-form. The supersymmetry spinor
therefore allows us to define forms which characterize the geometry
of the internal space.

How does this change when we include fluxes in the compactification?
The first thing to observe is the fluxes enrich (and complicate) the
supersymmetry variations, allowing more general backgrounds than
just Calabi-Yau spaces. Secondly, the gravitino variation implies
that the supersymmetry  spinor is no longer covariantly constant
with respect to the Levi-Civita connection, but is covariantly
constant with respect to a connection that involves the flux.
Schematically, \be \nabla_M \eta = 0 \,\,\longrightarrow \,\,(\nabla
+ {\rm flux})_M \eta = 0. \ee Although the manifold no longer has
reduced holonomy, it still has a reduced structure group $G\subset
\SU(4)$, and the deviation from special holonomy can be measured
using instrinsic torsion. In particular, $dJ$ and $d\Omega$ are
no-longer zero, and are sourced by the fluxes. To be more specific,
we will consider a general $N=1$ type II compactification. There are
two supersymmetry spinors, which can be written (in string frame):
\bea
\e^1 &=& \zeta_-\otimes \eta^1_+ + \cc,\non\\
\e^2 &=& \zeta_-\otimes \eta^2_+ + \cc,\label{eqn:susy_spinor_1}
\eea where $\zeta_\pm$ are $d=4$ Weyl spinors and $\eta^i_\pm$ are
internal Weyl spinors, with the sign denoting chirality. In this
notation complex conjugation corresponds to a flip in chirality. In
order to preserve $N=1$ supersymmetry, the two spinors $\eta^1$ and
$\eta^2$ need to be related. The type of relation characterizes the
internal geometry in terms of the structure group of the manifold.
In particular there are three obvious cases:
\begin{enumerate}
\item $\eta^1\propto \eta^2$ everywhere.  The structure group is at most $\SU(3)\subset \SU(4)$.
This class of solutions
typically come from large volume compactifications discussed in
\cite{Dasgupta:1999ss,Gukov:1999ya}, and are conformally Calabi-Yau.
\item $\eta^1 \perp \eta^2$ everywhere. The structure
 group of the internal manifold is reduced from $\SU(3)$ to $\SU(2)$
  and this imposes strict topological conditions on the internal manifold; for example, $\chi =0$.
The geometry is labeled ``static $\SU(2)$."

\item $\eta^1$ and $\eta^2$ interpolate between cases (1) and (2) at different
points on the internal space: there may be points where they are
parallel and other points where they are orthogonal. This is clearly
the most general type of solution and is called ``local $\SU(2)$."
\end{enumerate}



In our case the geometry will have local SU(2) structure, with a
structure group that includes the quantum $O(4,4,\Z)$ T-duality
group.\footnote{See Appendix~\ref{app:su2_susy}\ for a brief
overview of the necessary $SU(2)$ $G$-structure analysis.} The
novelty arises in the kinds of flux one can write down without
breaking supersymmetry. Our example is one that admits $(0,3)$ and
$(3,0)$ $G_3$-flux
along with non-trivial $F_1$ and $F_5$ fluxes.\footnote{It is
usually the case that the presence of $(0,3)$ or $(3,0)$ fluxes in
compact string solutions breaks supersymmetry. That is true for
models with a large volume limit. Here we relax that constraint.}

Although we derive the solution in the orbifold limit, the
background has moduli which give rise to a family of type IIB
solutions.
Our type IIB solution can also be lifted to M-theory where the
resulting flux is no longer necessarily $(2,2)$. This is in contrast
to the solutions typically studied which are based
on~\cite{Becker:1996gj}. Our new solutions are therefore examples of
the more general structures possible when a more general spinor
ans\"atze is used in solving the supergravity equations of motion.
We relegated some of the details required to demonstrate that the
solutions preserve supersymmetry and obey the equations of motion to
Appendix~\ref{app:su2_susy}.

\subsection{A non-geometric type IIB solution}

Our starting point is again the type I solution described in section
\ref{sect:typeI}\  with the choice of fluxes given
by~\C{eqn:fluxes_1}. Our parameterization of the flux and metric
imply there are isometries along the $(x_1,y_1)$ directions of
$T^4/\Z_2$, so we can T-dualize these directions using the Buscher
rules to construct a dual type IIB solution. For convenience, these
rules are summarized in Appendix~\ref{sect:t-duality}.

The $D9/O9$ system of type I becomes $D7/O7$-branes localized in the
fiber of $T^4/\Z_2$. Denoting the T-dualized fields by $\wt G$, $\wt
B$, $\wt \Phi$ and $\wt C_n$ we find the NS-NS background: \bea
\wt ds^2 &=& e^{-3\phi/4} \eta_{\mu\nu}dx^\mu dx^\nu + e^{3\phi/4}\left\{ \varpi \left[ dw d\bar w + dz_1 d\bz_1 \right] + dz_2d\bz_2\right\}\label{eqn:IIB_metric_1b}\\
e^{\wt\Phi_{IIB}} &=& \varpi\label{eqn:IIB_dilaton_1}, \\
\wt B_2&=& -A \varpi \bz_2 dz_1\w dw + \cc\label{eqn:IIB_bfield_1},
\eea where $dz_i = dx_i + i dy_i$, $dw = dw_1 + i dw_2$ and $\varpi$
is given in \C{varpi2}. Metrically the internal space $\cM_3$ is a
$T^4$ fibration over a $T^2$ base. We will explain below how to make
sense of this globally. The RR field content is \bea
\wt F_1 &=& -\star_{\,\PP^1}d\varpi^{-1} = \O(\alpha'^2),\label{eqn:f1_1}\\
\wt F_3 &=&
 -iA dw \wedge dz_1\w d\bz_2 + \cc + \O(\alpha'^2),\label{eqn:IIB_2form_1}\cr
\wt F_{5} &=& -A^2\varpi[ \star_{\,\PP^1} d(|z_2|^2) \w dz^1 \w
d\zbar^1 \w dw \w d\bar w] + {\rm
hodge~dual}+\O(\alpha'^2)\label{eqn:f5_1}, \eea where $\star_{T^2}$
denotes taking the Hodge dual  with respect to the unwarped metric
on the $\PP^1$ base. In the last line, we have taken the
ten-dimensional Hodge dual (so that $\wt F_5$ is self-dual). Note
that $dF_5=H_3\w F_3$, which is a good consistency check. The
spinors dualize as follows \bea
\wt \e_L &=&e^{-3\phi/16} \zeta_-\otimes\eta_+ + \cc, \non\\
\wt \e_R &=&e^{-3\phi/16} \zeta_-\otimes \left[c \eta_+ +
d\chi_+\right] + \cc\label{eqn:spinor_3}, \eea where $\eta_+$ and
$\chi_+$ are two orthogonal spinors defined on the unwarped internal
space. The coefficients are given by: \be c =\varpi\left( 4A^2
|z_2|^2-e^{3\phi/2}\right) \qquad d= 4 z_2 \varpi A e^{3\phi/4}. \ee
This is a compact type IIB vacuum with local $\SU(2)$ structure: at
generic points on the internal space, the spinors $\wt \e_L$ and
$\wt \e_R$ are neither orthogonal nor parallel, even at the level of
supergravity.

The solution is non-geometric in a fashion similar to the heterotic
solution we described in section~\ref{newhet}. In this case, the
internal space $\cM_3$ looks like a $T^4$ fibration over a
$T^2/\Z_2$ base. By including group elements from the $O(4,4,\Z)$
T-duality group (which can be thought of as coming from
compactifying type IIB on the $T^4$ fiber), we find the metric is globally
well-defined.


The internal space is therefore a fibration (after including the
non-geometric twists), with 7-branes localized in the $T^4$ fiber.
Although we deduced the presence of the 7-branes via T-duality, the
action of the non-geometric twists on the open string sector is
quite complex. As alluded to in the introduction, this is one of the
complications one must face when analyzing non-geometric
compactifications in type II string theory; M-theory and the
heterotic string do not have such issues.
\begin{figure}[htp]
\begin{center}\leavevmode
\xymatrix{ T^4 \ar[r] & \cM_3 \ar[d] \\
 & T^2/\Z_2 }
\end{center}
\caption{Fibration structure of the type IIB $\cM_3$
solution.}\label{fig:m6}
\end{figure}

We have checked that the background given above locally satisfies
the type IIB supersymmetry constraints. The details are rather
involved, and a summary is given in Appendix~\ref{app:su2_susy}.

As a further test, this solution can be lifted to M-theory (see the
following section) and the equations of motion checked. Explicitly
these are given by \bea R_{MN} -\half G_{MN} R &=&
\frac{1}{12}\left( G_{MPQR}G_{N}^{\,\,PQR} - \frac{1}{8} G_{MN}
G_{PQRS}G^{PQRS}\right). \eea The fluxes must also satisfy the
Bianchi identity, \be d*G_4 + \half G_4\wedge G_4 = \O(\ell_p^4).
\ee After performing the lift to M-theory,  one can see explicitly
that our solution satisfies Einstein's equations.

It is interesting to note that the typical supersymmetry constraints
used in the literature for studying type IIB flux compactifications
are that the $G_3$-flux must be imaginary self-dual, primitive and
$(2,1)$ with respect to the complex
structure~\cite{Dasgupta:1999ss}. Fluxes that do not obey these
constraints are typically thought to break supersymmetry. We have
constructed here a counter-example to this lore: a solution with
$G_3$-flux that is not $(2,1)$, consistent with a non-holomorphic
dilaton. Such solutions were first pointed out in
\cite{DallAgata:2004dk}\ and here we have constructed an example.
There are many ways to generalize this construction like starting
with both $\alpha$ and $\beta$ fluxes which would give N=1 models.

\subsection{Lift to M-theory}
We now lift the type IIB solution to M-theory in the usual way. This
is useful because we avoid the difficulties in defining orientifold
projections in non-geometric type IIB. So assume $\cM_3$ is the base
of a torus-fibered four-fold $\wt\cM_4$, with the complex structure of
the torus  determined by the type IIB string coupling. This is
depicted in figure~\ref{fig:fourfold}.
\begin{figure}[htp]
\begin{center}\leavevmode
\xymatrix{ T^2 \ar[r] & \wt\cM_4\ar[d] \\
 & \cM_3 \\
}
\end{center}
\caption{Fibration structure of the $\wt\cM_4$
solution.}\label{fig:fourfold}
\end{figure}
The three-fold $\cM_3$, whose fibration structure is illustrated in
figure~\ref{fig:m6}, is itself $T^4$-fibered (including
non-geometric twists). Therefore the M-theory solution itself only
makes sense using the appropriate U-duality group. Using the
standard relation between type IIB and M-theory, we can read off the
M-theory metric: \be ds_{11}^2 =e^{-\phi}\eta_{\mu\nu}dx^\mu dx^\nu
+ e^{\phi/2}(\varpi dw d\bar w + \varpi dz_1 d\bz_1 + dz_2 d\bz_2) +
e^{\phi/2}dvd\bar v.\label{eqn:m_metric_1b} \ee The coordinate $v$
parameterizes the torus fiber, and we absorb the volume into $v \sim
v + 2\pi R \sim v + 2\pi R \tau$. The complex structure of the torus
is given by the axio-dilaton of type IIB, \be \tau = \wt C_0 +
i(e^{3\phi/2} + 4A^2|z_2|^2), \ee where $d\wt C_0 = \wt F_1$ in
\C{eqn:f1_1}.

There is also the M-theory three-form $A_{MNP}$, which has internal
legs given by the type IIB two-forms. Explicitly, the three form has
one leg along the fiber and one along the base. It is determined in
terms of type IIB fluxes, \bea
B_{\mu\nu} &\longleftrightarrow& A_{\mu\nu v^1},\\
C_{\mu\nu} &\longleftrightarrow& A_{\mu\nu v^2}, \eea with
$A_{MNP}=0$ otherwise. There is also the type IIB five-form field
strength with four legs in space-time specified by~\C{eqn:f5_1}.
This lifts to a space-time filling four-form field strength
$G_{012a}$ where $G=dA$.

Let us examine the global behavior. Under the periodicities $z_2\sim
z_2+2\pi\sim z_2+2\pi \tau$ the metric and complex structure are
defined only up to U-duality transformations. We have arrived at
solution of M-theory that is locally geometric, but globally
non-geometric, requiring patching by U-duality. This is a U-fold as
originally sought in~\cite{Kumar:1996zx}\ but of a quite different
local form.

A short note on dualizing the spinor. It is possible to show that
the type IIB spinor lifts to a Majorana spinor in M-theory of the
form, \bea \ve = e^{-5\phi/4}\psi \otimes (\xi_1 +
\xi_2)\label{eqn:mtheory_spinor_1}. \eea Here $\xi_1$ and $\xi_2$
are $d=8$ Majorana spinors which have chiral components: \be \xi_i =
\xi_i^+ + \xi_i^-, \ee with $\xi_i^\pm$ Majorana-Weyl spinors.
Because of the zeroes in $c$ and $d$ defined in
\C{eqn:su2_spinor_2}, these chiral components will also have zeroes.
This is a background of the type discussed by
Tsimpis~\cite{Tsimpis:2005kj}, in which we preserve $N=2$ in $d=3$.
This more general spinor ansatz is the reason one can have more
general flux configurations (i.e. not necessarily $(2,2)$ fluxes)
without breaking supersymmetry.

\newpage

\appendix

\section{\bf  Weierstrass models for maximal gauge symmetry}
\label{app:Weierstrass}

In this appendix, we summarize the derivation of eqs.~\eqref{eq:MV}\ and~\C{eq:AMbis}, following \cite{FCY2,MR1416960,Aspinwall:1996vc,instK3}.
To derive eq.~\eqref{eq:MV},
we use Weierstrass form: in order to get fibers of type $II^*$ at both
$\sigma=0$ and $\sigma=\infty$, Kodaira's table implies that
the coefficient
of $X$ must have zeros of order $4$ at both $0$ and $\infty$. Since
this coefficient
has degree $8$, it therefore takes the form $a \sigma^4$ in affine coordinates.
Similarly, the coefficient of $X^0$ must have zeros of order $5$ at
both $0$ and $\infty$, with overall degree $12$; thus, it must take the
form $b \sigma^5+c \sigma^6+d\sigma^7$.

To derive eq.~\eqref{eq:AMbis}, we start by imposing a $\mathbb Z_2$
subgroup of the Mordell--Weil
group.  (We do this because the desired gauge group
$\overline{Z}_{\operatorname{Spin}{32}/\mathbb Z_2}$ has fundamental
group $\mathbb Z_2$, which implies that the torsion part of
the Mordell--Weil group should be $\mathbb Z_2$ \cite{pioneG}.)
Having a $\mathbb Z_2$ subgroup of the Mordell--Weil group
 means that there should be a point of order $2$ on the elliptic curve,
and by a translation in the $(x,y)$ plane we can move this point to $(0,0)$.
For $(0,0)$ to be a point of order two, we need the tangent line of the
elliptic curve at this point to be vertical; this implies that the equation
takes the form
\begin{equation}
 y^2 = x^3 + p_4(s) x^2 + \varepsilon_8(s) x ,
\end{equation}
where $p_4(s)$ and $\varepsilon_8(s)$ are polynomials of degree $4$ and
$8$, respectively.  Using the substitution $x=\bar{x}-p_4(s)/3$, we can
restore this to Weierstrass form:
\begin{equation}
y^2 = \bar{x}^3 + \left(\varepsilon_8(s)-\frac13 p_4(s)^2\right)\bar{x} +
p_4(s)\left(\frac2{27}p_4(s)^2-\frac13\varepsilon_8(s)\right),
\end{equation}
which allows us to compute the discriminant:
\be
\begin{aligned} \Delta &= 4\left(\varepsilon_8(s)-\frac13 p_4(s)^2\right)^3
+ 27 p_4(s)^2 \left(\frac2{27}p_4(s)^2-\frac13\varepsilon_8(s)\right)^2\\
&= \varepsilon_8(s)^2\left( 4\varepsilon_8(s) - p_4(s)^2 \right).
\end{aligned}
\ee

To get a fiber of type $I_{12}^*$ at $s=\infty$, $f$, $g$, and $\Delta$
must have zeros of order $2$, $3$, and $18$ (respectively) at $s=\infty$.
Thus, with respect to the affine coordinate $s$, we must have
\be
\begin{aligned}
\deg \left(\varepsilon_8(s)-\frac13 p_4(s)^2\right) &= 8-2=6\\
\deg  \left(p_4(s)\left(\frac2{27}p_4(s)^2-\frac13\varepsilon_8(s)\right)\right)
 &= 12-3=9
\end{aligned}
\ee
and
\be
 \deg \Delta = \deg \varepsilon_8(s)^2\left(4\varepsilon_8(s)-p_4(s)^2\right) = 24-18=6.
\ee

From this data, we argue as follows.  First,
if $\deg p_4(s)=4$, then there must
be cancellation between leading terms of $p_4(s)^2$ and $\varepsilon_8(s)^2$ to get
lower degree for
both $\varepsilon_8(s)-\frac13 p_4(s)^2$ and $\frac2{27}p_4(s)^2-\frac13\varepsilon_8(s)$.
But since those linear combinations are not proportional to each other,
it is not possible to acheive both cancellations.  Thus, $\deg p_4(s)\le 3$.
To get the correct reductions in degree, it is easy to see that
also $\deg \varepsilon_8(s)\le6$.  But now if $\deg p_4(s)<3$, the second combination
would have its degree reduced below $9$.  Thus $\deg p_4(s)=3$. In
eq.~\eqref{eq:AMbis} and also in Appendix~\ref{CDrefinement},
we refer to this cubic polynomial as
$p(s)=p_0s^3+p_1s^2+p_2s+p_3$.

It then follows that $\deg (4\varepsilon_8(s)-p_4(s)^2)=6$, and so to achieve the
correct order of vanishing of the discriminant, the degree of $\varepsilon_8(s)$
must be ${}\le0$, i.e., $\varepsilon_8(s)$ must be constant. In
eq.~\eqref{eq:AMbis} and also in Appendix~\ref{CDrefinement},
we simply refer to this constant as $\varepsilon$.

\section{\bf An explicit Shioda--Inose structure}
\label{CDrefinement}

In this appendix, we describe the explicit Shioda--Inose structure
found by Clingher and Doran \cite{math.AG/0602146}, and make it more
precise.  Our first step is to make explicit the involution on the
$K3$ surface
$\overline{Z}_{\operatorname{Spin}{32}/\mathbb Z_2}$,
and to compute the quotient by that involution.

The involution on $\overline{Z}_{\operatorname{Spin}{32}/\mathbb Z_2}$
is
induced by translation by the point
of order $2$.  To work this out geometrically, we start with an
arbitrary point $(x_0,y_0)$ on the elliptic curve
and connect it by a line to $(0,0)$;
this line has equation $y=(y_0/x_0)x$.  Substituting in, we find
\be
 (y_0^2/x_0^2)x^2=x^3 + p(s) x^2 + \varepsilon  x
\ee
or
\be
\begin{aligned}
 0&=x^3 + \left(p(s)-y_0^2/x_0^2\right) x^2 + \varepsilon  x\\
&= x(x^2 + \left(p(s)-y_0^2/x_0^2\right) x + \varepsilon )\\
&=x(x-x_0)(x-\varepsilon /x_0)
\end{aligned}
\ee
since
\be
 -x_0-\varepsilon /x_0=p(s)x-y_0^2/x_0^2.
\ee
The third point of intersection with the line is therefore
\be
 (\varepsilon /x_0, \varepsilon y_0/x_0^2).
\ee
The translation by $(0,0)$ yields the point with the same $x$ value, but
the negative of the $y$ value; that is, our automorphism is:
\be
 (x,y) \mapsto (\varepsilon /x, -\varepsilon y/x^2).
\ee

The quotient can be described by introducing invariants
\be
\begin{aligned}
\xi &= x+ \frac{\varepsilon}x \\
\eta &= y - \frac{\varepsilon y}{x^2}
\end{aligned}
\ee
and observe that our equation can be written $y^2=x^3+p(s)x^2+\varepsilon x
= x^2(\xi+p(s)).$  Then
\be
\begin{aligned}
\eta^2 &= y^2 \left(1-\frac{\varepsilon y}{x^2}\right)^2\\
&=x^2(\xi + p(s))\left(1-\frac{\varepsilon y}{x^2}\right)^2\\
&= (\xi +p(s))\left(x-\frac{\varepsilon y}{x}\right)^2\\
&= (\xi +p(s))(\xi^2-4\varepsilon).
\end{aligned}
\ee
This is the equation of the quotiented surface.  Its discriminant is
\begin{equation}\label{eq:quotdisc}
 -16\varepsilon(p(s)^2-4\varepsilon)^2,
\end{equation}
which has roots at $s=\infty$ and at the roots of $p(s)\pm2\sqrt\varepsilon$
(the latter are all double roots).

Clingher and Doran \cite{math.AG/0602146} start with the Kummer
surface of the product $E\times F$, where $E$ is the double cover
of $\mathbb{CP}^1$
with branch points $\{0, 1, \alpha, \infty\}$
and
$F$ is the double cover of $\mathbb{CP}^1$ with branch points
$\{0, 1, \beta, \infty\}$.
Clingher and Doran use the analysis of Oguiso \cite{MR1013073}
to locate the elliptic fibration on the Kummer surface
$\operatorname{Km}(E\times F)$ which corresponds to unbroken
$\operatorname{Spin}{32}/\mathbb Z_2$, and find that the singular fibers of type
$I_2$ of that fibration are located at $6$ specific values of the
parameter $t$ of the fibration,
divided into two groups of three \cite[eq. 59]{math.AG/0602146}:
\be
 \left\{ 1, \frac 1\alpha, \frac 1\beta, \frac 1{\alpha\beta}, \frac{\alpha\beta+1}{\alpha\beta},
\frac{\alpha+\beta}{\alpha\beta}\right\}
= \left\{ 1, \frac1{\alpha\beta}, \frac{\alpha+\beta}{\alpha\beta} \right\}
\cup
\left\{ \frac 1\alpha, \frac 1\beta, \frac{\alpha\beta+1}{\alpha\beta}\right\}.
\ee
By rescaling the parameter $t$ to $s=(\alpha\beta)t$, we can scale all of these singular values
by $\alpha\beta$, giving the values
\be
 \left\{ \alpha\beta, \beta, \alpha, 1, {\alpha\beta+1},
{\alpha+\beta}\right\}
= \left\{ \alpha\beta, 1, {\alpha+\beta} \right\}
\cup
\left\{  \beta,  \alpha, {\alpha\beta+1}\right\}.
\ee
The monic polynomials which vanish on these two sets
\be
\begin{aligned}
D_1(s)&=(s-\alpha\beta)(s-1)(s-\alpha-\beta)\\
D_2(s)&=(s-\beta)(s-\alpha)(s-\alpha\beta-1)
\end{aligned}
\ee
 then have the remarkable property that their difference
\be
D_1(s)-D_2(s)=\alpha\beta(\alpha-1)(\beta-1)
\ee
 is a constant and, in particular, is independent of
$s$.  Thus, for some monic polynomial $q(s)$, the polynomials $D_j(s)$ can
be written as $q(s)\pm\frac12\alpha\beta(\alpha-1)(\beta-1)$.

Since the only singular fibers of the elliptic fibration away from $s=\infty$
are the $I_2$ fibers which are located at the roots of $D_j(s)$, and at each
of which  the discriminant
 has a double zero, the discriminant must be (up to a multiplicative
constant):
\be
 D_1(s)^2D_2(s)^2 = (q(s)+\frac12 C)^2(q(s)-\frac12 C)^2
= (q(s)^2-\frac14{C^2})^2,
\ee
where $C=\alpha\beta(\alpha-1)(\beta-1)$.  This is precisely the form which
we derived in eq.~\eqref{eq:quotdisc}, if we identify $C^2/4$ with
$4\varepsilon$ and $q(s)$ with $p(s)$.

In other words, in our quotient
$\Z_{\operatorname{Spin}(32)/\mathbb Z_2}/\iota_{\operatorname{Spin}(32)/\mathbb Z_2}$,
the roots of $p(s)\pm2\sqrt\varepsilon$ are the two sets
\be
 \left\{ \alpha\beta, 1, {\alpha+\beta} \right\} \text{ and }
\left\{  \beta,  \alpha, {\alpha\beta+1}\right\}.
\ee
It is then easy to derive the formulas:
\be
p(s)=s^3-(\alpha+1)(\beta+1)s^2
+((\alpha+\beta)(1+\alpha\beta)+\alpha\beta)s - \frac12\alpha\beta(\alpha+1)(\beta+1)
\ee
\be
\varepsilon=\frac1{16}\alpha^2\beta^2(\alpha-1)^2(\beta-1)^2,
\ee
since
\be
\frac{\alpha\beta(1+\alpha\beta)+\alpha\beta(\alpha+\beta)}2
=\frac{\alpha\beta(\alpha+1)(\beta+1)}2
\ee
and
\be
\frac{\alpha\beta(1+\alpha\beta)-\alpha\beta(\alpha+\beta)}2
=\frac{\alpha\beta(\alpha-1)(\beta-1)}2.
\ee

We now wish to generalize this relation to a pair of elliptic curves
for which the equations have been given but not the set of branch points.
To this end, let
\be
 v^2=(u-\alpha_1)(u-\alpha_2)(u-\alpha_3)
=u^3+\lambda_1u^2+\lambda_2u+\lambda_3
\ee
and
\be
 w^2=(z-\beta_1)(z-\beta_2)(z-\beta_3)=z^3+\mu_1z^2+\mu_2z+\mu_3.
\ee
define $E$ and $F$.  We claim that in this case, the two triples of
roots of $p(s)\pm2\sqrt\varepsilon$ will be given by
\be
\{
\alpha_1\beta_1+\alpha_2\beta_2+\alpha_3\beta_3,
\alpha_2\beta_1+\alpha_3\beta_2+\alpha_1\beta_3,
\alpha_3\beta_1+\alpha_1\beta_2+\alpha_2\beta_3
\}
\ee
and
\be
\{
\alpha_1\beta_1+\alpha_3\beta_2+\alpha_2\beta_3,
\alpha_2\beta_1+\alpha_1\beta_2+\alpha_3\beta_3,
\alpha_3\beta_1+\alpha_2\beta_2+\alpha_1\beta_3
\}
\ee
To verify that these are the same roots as before,
we can set $\alpha_1=\beta_1=0$,
$\alpha_2=\beta_2=1$, $\alpha_3=\alpha$ and $\beta_3=\beta$; then these two triples
reduce to the previous case.  Since the pair of triples is invariant
under the action of the symmetric group on either set of roots,
the sets are the same.

Now, however, the polynomials $p(s)\pm2\sqrt\varepsilon$ can be determined by
a computation with the elementary symmetric functions ($\lambda$'s and $\mu$'s)
of the roots ($\alpha$'s and $\beta$'s).  The result is:
\be
\begin{gathered}
 p_1 = -\lambda_1\mu_1
\\
 p_2 = \lambda_1^2\mu_2   + \lambda_2\mu_1^2 -3\lambda_2\mu_2
\\
 p_3 = -\lambda_1^3\mu_3 -\lambda_3\mu_1^3 -\frac12\lambda_1\lambda_2\mu_1\mu_2+\frac92\lambda_1\lambda_2\mu_3+\frac92\lambda_3\mu_1\mu_2
-\frac{27}2\lambda_3\mu_3
\\
 \varepsilon = \frac1{16}\operatorname{disc}_u(u^3{+}\lambda_1u^2{+}\lambda_2u{+}\lambda_3) \operatorname{disc}_z(z^3{+}\mu_1z^2{+}\mu_2z{+}\mu_3).
\end{gathered}
\ee
By completing the cube (in $u$, in $z$, and in $s$),
we can set $\lambda_1=\mu_1=p_1=0$, leaving
\be
\begin{gathered}
 p_2 =  - 3\lambda_2\mu_2
\\
 p_3 =
-\frac{27}2\lambda_3\mu_3
\\
 \varepsilon = \frac1{16}(4\lambda_2^3+27\lambda_3^2)(4\mu_2^3+27\mu_3^2).
\end{gathered}
\ee

\section{Some notation and useful relations}
\label{appendixnotation}

In this appendix, we summarize some notation and  relations of use
in the construction of the explicit solutions of
sections~\ref{hettorsion}-\ref{iibm}.

We take coordinates $(x^1,y^1,x^2,y^2)$ for $K3$, while for
$\wt{K3}$ we use $(w^1,w^2,v^1,v^2)$. Both surfaces are assumed to
be elliptically fibered. We use Roman indices to denote the
following: $m,n=x^1,\ldots,y^2$ the coordinates for the total space
$K3$; $p,q=x^2,y^2$ the $\PP^1$ base of the $K3$; $i,j=x^1,y^1$ the
elliptic fiber of $K3$. The indices $a,b$ are tangent space indices
for the internal space. We also use the complex combinations
$dz^\alpha = dx^\alpha + i dy^\alpha$, for $\alpha=1,2$ and $dw =
dw^1 + i dw^2$.

The function $\varpi$ appears often: \be \varpi =
\frac{1}{e^{3\phi/2} + |B_{w^1}|^2 + |B_{w^2}|^2}.\label{varpi3} \ee
where  $|B_{w^i}|^2 = (B_{w^i x^1})^2+(B_{w^i y^1})^2$ and $\varpi$
always satisfies $\varpi=1+\O(\alpha'^2)$.

The relation between complex and real components for vectors and
one-forms are summarized by: \be C_w = \half ( C_{w^1} - i C_{w^2}),
\quad C_{w^1} = C_w + C_\bw, \quad C_{w^2} = i(C_w - C_\bw). \ee
Hodge duality on an $m$-dimensional manifold is defined as \be \star
F_p = \frac{\sqrt{G}}{p! (m-p)!}
\ve^{\nu_1\ldots\nu_p}_{~~~~~~\mu_{p+1}\ldots \mu_{m}} F_{\nu_1
\ldots \nu_p}dx^{\mu_{p+1}}\w\ldots\w dx^{\mu_{m}}. \ee When taking
Hodge dual it is useful to differentiate between warped and unwarped
metrics. In particular, $\ve$ will be with respect to the warped
metric, while $\e$ is unwarped. The Hodge dual squares to \bea \star
\star F_p &=& (-1)^{p(m-p)} F_p, \quad {\rm
for~a~Riemannian~space},\cr \star \star F_p &=& (-1)^{p(m-p)+1}
F_p,\quad {\rm for~a~Lorentzian~space}. \eea The adjoint
differential operator is defined as \bea d^\dagger &=&
(-1)^{(p+1)m+1} \star d \star \quad {\rm for~a~Riemannian~space},\cr
d^\dagger &=& (-1)^{(p+1)m} \star d \star \quad {\rm
for~a~Lorentzian~space}, \eea while the Laplacian $\Box = d
d^\dagger + d^\dagger d$.

The RR field strengths are defined by $F_{n+1} = d C_{n} + H_3 \w
C_{n-3}$, obeying $$\star_{10}F_n = (-1)^{\lfloor n/2 \rfloor}
F_{10-n}.$$

\section{The T-duality rules}
\label{sect:t-duality}

In this appendix, we summarize the Buscher rules which are extensively used in
constructing the new metrics.

\subsection{Metric and fluxes}
The Buscher rules determine the value of the metric and B-field.
They are given by \cite{Johnson:2000ch}, \bea
\wt G_{99} &=& G_{99}^{-1}, \non\non\\
\wt G_{i9} &=& G_{99}\inv B_{i9}, \non\non\\
\wt B_{i9} &=& G_{99}\inv G_{i9}, \non\non\\
\wt G_{ij} &=& G_{ij} - G_{99}\inv(G_{9i}G_{9j} -
B_{9i}B_{9j}),\non\non\\
\wt B_{ij} &=& B_{ij} - G_{99}\inv(G_{9i}B_{9j} - B_{9i}G_{9j}),
\non\non\\
2\wt\phi &=& 2\phi - \ln G_{99}, \eea where $i=0,\ldots,8$ and $X^9$
is the isometry direction along which we T-dualize. The dilaton
becomes \be e^{\Phi'} = e^{\Phi} \left( \frac{\det \wt G}{\det G}
\right)^{1/4}. \label{eqn:buscher_dilaton} \ee The RR fluxes dualize
as follows\cite{Bergshoeff:1995as, Johnson:2000ch}: \bea \wt
C^{(n)}_{\mu\ldots\nu\alpha 9} &=& C^{(n-1)}_{\mu\ldots\nu\alpha} -
(n-1)G_{99}\inv C^{(n-1)}_{[\mu\ldots\nu|9|} G_{\alpha] 9},\non\\
\wt C^{(n)}_{\mu\ldots\nu\alpha\beta} &=&
C^{(n+1)}_{\mu\ldots\nu\alpha\beta 9} + n
C^{(n-1)}_{[\mu\ldots\nu\alpha}B_{\beta] 9} + n(n-1) G_{99}\inv
C^{(n-1)}_{[\mu\ldots\nu|9|}B_{\alpha|9|}G_{\beta] 9},
\label{eqn:buscher_rr_1} \eea where $C$ denotes the original fluxes
and $\wt C$ the T-dualized fluxes. Here $\mu,\nu,\alpha\ldots \ne
9$.

\subsection{Spinors under T-duality}
The spinors dualize according to the rules written down by
Hassan\cite{Hassan:1999bv}. For T-duality of the supergravity
spinors $\e_{1,2}$ this is a simple generalization of the flat space
T-duality rules: \bea
\e_L &\rightarrow& \e_L,\non\\
\e_R &\rightarrow& \beta_9 \e_R. \eea In flat space, the space-time
indices coincide with tangent space indices and $\beta_9 =
\Gamma\Gamma_9$ as usual. In a curved background, one simply
generalizes $\beta_9$: \be \beta_9 =
\sqrt{G_{99}\inv}\Gamma\Gamma_9, \ee where $G_{MN}$ the original
metric, with $M,N$ space-time indices. The gamma matrices satisfy
$\{\Gamma^M,\Gamma^N\} = 2 G^{MN}$, $\{\Gamma,\Gamma^M\} = 0$ and
$\Gamma^2 = 1$. Further, $\Gamma_9 = G_{9M}\Gamma^M$, or in terms of
Lorentz frame indices $a,b,\ldots$ we have $\Gamma_9 =
e_{9a}\Gamma^a$, where $G_{MN} = e^{\,\,\,a}_M \eta_{ab}
e^b_{\,\,\,N}$ and $e_{Ma} = G_{MN} e^N_{\,\,\,a}$. The
normalization is determined by $\beta^2 = e^{i\pi F_R}$.

The vielbein transforms under T-duality as follows \be \wt e^M_a =
Q^M_{\,\,N} e^N_a \ee where \be
Q^M_{\,\,N} = \begin{pmatrix} G_{xx} & (G + B)_{xa} \\
0 & 1
\end{pmatrix}.
\ee The left-moving vielbein is invariant, while the right-moving
vielbein transforms in the following way \be e^M_a \rightarrow
Q^M_{\,\,N} \, e^N_a. \ee

\section{Type IIB supergravity with $\SU(2)$ structure}
\label{app:su2_susy}  In this section, the constraints supersymmetry
imposes on the fluxes and geometry are reviewed for a general
$\SU(2)$ structure spinor ansatz. Such an analysis was first
performed in~\cite{DallAgata:2004dk}, and we review that work here
because
the solutions derived in section $4$ are examples of this type.

 The
spinor basis used in~\cite{DallAgata:2004dk}\ is inconvenient for
our purposes, and so we will rederive the pertinent results  in a
more convenient basis and notation.


\subsection{Type IIB supersymmetry}
In this paper we are interested in $d=4$ compactifications
preserving $N=1$ supersymmetry in space-time. The most general
metric with Minkowski space-time takes the form \be ds^2 =
e^{-3\phi/4} \eta_{\mu\nu}dx^\mu dx^\nu + e^{3\phi/4} g_{ab}dy^a
dy^b,\label{eqn:iib_metric_1}\non \ee where the internal metric
$g_{ab}$ will in general have $\SU(2)$ structure. We formulate the
type IIB supergravity variations in Einstein frame where the
$\SL(2,\R)/\U(1)$ symmetry of type IIB supergravity is manifest. In
section~\ref{appendixsolve}, we will switch to string frame which is
convenient for performing string dualities.

The only non-trivial variations are those of the dilatino and
gravitino: \bea \delta \lambda &=& \frac{i}{\kappa} \wt\Gamma^M P_M
\ve^* - \frac{i}{4} \wt G
\ve, \label{eqn:sugra_1}\\
\delta \Psi_M &=& \frac{1}{\kappa}\wt D_M \ve + \frac{i}{480} \wt
\Gamma^{M^1\ldots M^5}F_{M_1\ldots M_5} \wt \Gamma_M \ve -
\frac{1}{16} \wt \Gamma_M \wt G \ve^* - \frac{1}{8} \wt G \wt
\Gamma_M \ve^*.\label{eqn:sugra_2} \eea The supersymmetry spinor
$\ve$ is a complex $d=10$ Weyl spinor, and the tilde denotes gamma
matrices that are defined with respect to the warped metric. The
field content of type IIB consists of a three-form $G_3$,
axio-dilaton $\tau$ and self-dual five-form $F_5$. Here $\wt G =
\frac{1}{6} G_{MNP}\wt \Gamma^{MNP}$. The derivative of the dilaton
and $\U(1)$ connection are given by \bea &&P_M = f^2 \del_M B, \quad
Q_M = f^2 \Im (B\del_M B^*), \quad {\rm
with}\non\\
&& B = \frac{1+i\tau}{1-i\tau}, \quad f^{-2} = 1- BB^*, \quad \tau =
C_0 + ie^{-\Phi}. \eea To preserve $d=4$ Poincare invariance, we
require all fields to depend only on internal coordinates, $G_3$ to
have only internal legs, and $F_5$ to be space-time filling viz. \be
F_{5} = (1 + \star)dx^0\w dx^1 \w dx^2 \w dx^3  \w dh, \ee with
$\star$ the $d=10$ Hodge dual and $h=h(y)$ an arbitrary scalar
defined on the internal manifold. The relation to the usual string
frame quantities $F_3,H_3$ is given by \bea \kappa G_3 = i
e^{i\theta}\frac{ F^s_3 - \tau H^s_3}{\tau_2^{1/2}}, \quad {\rm
with} \quad e^{i\theta} = \left(\frac{1 + i \tau^*}{1 -
i\tau}\right)^{1/2}. \eea The five-form is rescaled \be 4\kappa F_5
= F_5^s, \ee where $s$ denotes string frame quantities. Newton's
constant is given by $2\kappa^2 = (2\pi)^7g^2 \alpha'^2$. The metric
and spinor are also rescaled \be G_{MN} = e^{-\Phi/2} G^s_{MN},
\qquad \ve = e^{-\Phi/8} (\e_1 + i\e_2). \ee

\subsection{SU(2) structure}
Let us now turn to the spinor analysis of the type IIB supersymmetry
variations. The most general spinor ansatz preserving $N=1$
space-time supersymmetry is given in \C{eqn:susy_spinor_1}, which in
Einstein frame, takes the form \be \ve = \zeta_- \otimes \eta^1_+ +
\zeta_+\otimes \eta^2_-,\label{eqn:su2_spinor_1} \ee where
$\zeta_\pm$ are the $d=4$ space-time supersymmetry spinors while
$\eta^{1,2}_\pm$ are complex $d=6$ Weyl spinors, with $\pm$ denoting
chirality. The presence of $\SU(2)$ structure implies that there are
two orthogonal well-defined spinors $\eta_+$ and $\chi_+$. For a
supersymmetric solution preserving local $\SU(2)$ structure, we can
expand $\eta^{1,2}$ as \bea
\eta^1_+ &=& a\eta_+ + b\chi_+,\\
\eta^2_+ &=& c\eta_+ + d\chi_+.\label{eqn:su2_spinor_2} \eea

In $\SU(2)$ structure manifolds the spinors $\eta_+$ and $\chi_+$
are used to form spinor bilinears. These bilinears give a unique
characterization of the geometry. The spinors $\eta_\pm$ and
$\chi_\pm$ can be normalized so that $\eta_\pm^\dagger \eta_\pm =
\chi_\pm^\dagger\chi_\pm = 1$. They are related by \be \chi_+ =
\half w_m\gamma^m \eta_-, \ee where $\{ \gamma^m, \gamma^n\}=
2g^{mn}$ refer to the unwarped metric. We can now form the spinor
bilinears: \bea w_m = \eta_+^T \gamma_m \chi_+ , &\quad& J_{mn} =
-i\eta_-^T
\gamma_{mn}\eta_+,\non\\
K_{mn} = \eta_+^T \gamma_{mn}\chi_-, &\quad& \Omega_{npq} = -
\eta_+^T \gamma_{npq} \eta_+.\label{eqn:su2_invariants_1} \eea These
bilinears encode information about the local geometry. By taking
their exterior derivative one gets expressions for the intrinsic
torsion modules. The intrinsic torsion allows one to read off
various metric properties -- for example, whether the metric is
complex, K\"ahler, Calabi-Yau etc. Note that we have defined the
forms~\C{eqn:su2_invariants_1}\ with respect to the unwarped metric.
Further, we have chosen a complex structure $J$ defined by $\eta_+$
and corresponding $(3,0)$-form $\Omega$. With $\SU(2)$ structure
this is not the only choice: there is in fact, a $\U(1)$ of choices
of complex structure given by \be J = -i\beta_-^T \gamma_{mn}
\beta_+,\non \ee where \be \beta_+ =  \eta_+\cos\phi +
\chi_+\sin\phi.\non \ee The choice of $\phi$ will not affect whether
the solution is supersymmetric or not, but does affect its
interpretation. For example, the integrability of the complex
structure depends on the choice of $\phi$\cite{DallAgata:2004dk}.

With respect to the choice complex structure in
\C{eqn:su2_invariants_1} above, the $1$-form $w$ is holomorphic, as
is the $2$-form $K$ with $K= J^2 + i J^3$. Further, \be J = J^1 +
\frac{i}{2} w \w \bar w, \quad \Omega = K \w w.\non \ee

\subsection{Solving type IIB supergravity with $SU(2)$ structure}
\label{appendixsolve}

We will now rewrite the type IIB supersymmetry conditions
\C{eqn:sugra_1}-\C{eqn:sugra_2} in terms of the forms $w_m,J_{mn},$
and $K_{mn}$ defined in the previous section following
\cite{DallAgata:2004dk}. As a first step we decompose the fluxes
into $SU(2)$ forms.

\medskip\noindent {\bf Dilaton}:
\be P_m = p_1 w_m + p_2 \bar w_{m} + \Pi_m,\label{eqn:su2_dilaton_1}
\ee where $\Pi_m$ is a real one-form with $w \lrcorner \Pi=0$.

\medskip\medskip\noindent {\bf Three-form flux}:

\medskip
\noindent We decompose a general $3$-form flux in terms of the
complex structure given above, \bea G_3 &=& g_{(3,0)} K \w w +
g_{(2,1)} K \w \bar w + \wt g_{(2,1)} J^1
\w w + J^1 \w V^1 + w \w \bar w \w V^2 + w \w T^1 \non\\
&&\qquad + \bar w \w T^2 + g_{(1,2)} \bar K \w w + \wt g_{(1,2)} J^1
\w \bar w + g_{(0,3)} \bar K \w \bar w.\label{eqn:su2_g3_1} \eea
This is the most general expansion of $G_3$ in terms of SU(2)
modules. Here $V^i$ and $T^i$ are $1$-forms and $2$-forms,
respectively, orthogonal to $K$ and $w$: \be V^i \lrcorner w = 0,
\quad T^i \lrcorner w = 0.\non \ee The components $g_{(2,1)}$ and
$g_{(1,2)}$ are primitive while $g_{(2,1)}$ and $g_{(1,2)}$ are
non-primitive. Further, combining $V^1$ and $V^2$, the primitive and
non-primitive components can be made explicit viz. \be J^1 \w V_1 +
w\w \bar w \w V_2 =  \half (J^1 - \frac{i}{2} w \w \bar w) \w (V_1 +
2i V_2) + \half J \w (V_1 - 2iV_2). \ee
 The imaginary self-dual (ISD) limit corresponds to
\be g_{30} = g_{12} = \wt g_{21} = 0, \quad T^2 = 0, \quad
(1+iJ)(V^1+2iV^2) = (1-iJ)(V^1-2iV^2) = 0. \ee

\subsubsection*{Five-form flux and the warp factor:}
Lastly, we may similarly expand the warp factor: \bea
&&\del_n(\log e^{3\phi/2}) = \sigma w_n + \bar\sigma \bar w_n + \Sigma_n\non,\\
&&(\del_n h) = \theta w_n + \bar \theta \bar w_n + H_n.\non \eea

\subsection{The supersymmetry variations}

\subsubsection*{Dilatino:}
First we use the metric ansatz~\C{eqn:iib_metric_1}\ together with
the spinor ansatz~\C{eqn:su2_spinor_1}\ and plug it into the
dilatino variation \C{eqn:sugra_1} giving: \bea &&e^{-3\phi/8}
\frac{i}{\kappa} (\gamma^5\otimes \gamma^nP_n)
\left[\zeta_-\otimes(a^* \eta_+ + b^* \chi_+) + \zeta_+
\otimes(c^*\eta_- + d^* \chi_-\right] \non\\
&=& e^{-9\phi/8} \frac{1}{24}(\gamma^5 \otimes \gamma^{npq})
\left[\zeta_+ \otimes (a\eta_- + b\chi_-) + \zeta_-\otimes (c\eta_+
+ d\chi_+)\right]G_{npq}.\non \eea The gamma matrices $\gamma^a$ and
$\gamma^\mu$ are defined with respect to the unwarped metric, which
results in the warp factor appearing. Because $\zeta_+$ and
$\zeta_-$ are independent, the variations proportional to these
spinors must vanish. This gives \bea &&\gamma^n P_n (a^*\eta_+ +
b^*\chi_+) - \frac{e^{-3\phi/4}\kappa}{24}
\gamma^{npq}G_{npq}(c\eta_+ + d\chi_+) = 0,\non\label{eqn:dilatino_1}\\
&& \frac{e^{-3\phi/4}\kappa}{24} \gamma^{npq}G_{npq}(a\eta_- +
b\chi_-) -\gamma^n P_n(c^*\eta_- + d^*\chi_-) =
0.\non\label{eqn:dilatino_2} \eea A complete basis is specified by
$\eta_\pm$ and $\gamma^n\eta_\pm$. We contract with these to rewrite
the dilatino variation in terms of SU(2) invariants. This gives the
equations \bea 2e^{3\phi/4} p_1 d^* &=& -\kappa \left[2a g_{(3,0)} -
ib \wt
g_{(2,1)}\right],\label{eqn:dil_susy_2} \\
2 e^{3\phi/4} p_2b^* &=& -\kappa \left(2c\,g_{(0,3)}  + id\, \wt
g_{(1,2)}\right),\label{eqn:dil_susy_1}\\
2a^* c g_{(0,3)} &=& -i(b^* c + a^* d) \wt g_{(1,2)} - 2b^* d
g_{(2,1)},\label{eqn:dil_susy_5}\\
2a c^* g_{(3,0)} &=& \wt g_{(2,1)} i( ad^* + bc^*) - 2bd^*
g_{(1,2)}.\label{eqn:dil_susy_6} \eea From contracting with
$\gamma^k\eta_\pm$, we find \bea && (g+iJ)^{kn}\Pi_n c^* +
K^{kn}\Pi_n d^* =  \cr && -i\frac{\kappa e^{-3\phi/4}}{4}
\left[(g-iJ)^{kn}(V_1 - 2iV_2)_n a- bK^{kn}(V^1 + 2iV^2)_n
\right],\label{eqn:dil_susy_7}\\ &&(g+iJ)^{kn}\Pi_n a^* + \bar
K^{kn}\Pi_n b^* = \cr && i\frac{\kappa e^{-3\phi/4}}{4} \left[c
(g+iJ)^{kn}(V^1-2iV^2)_n - d \bar K^{kn}
(V^1+2iV^2)_n\right].\label{eqn:dil_susy_8} \eea The terms
$V^1+2iV^2$ and $V^1-2i V^2$ are part of the primitive and
non-primitive components of the flux, respectively. These terms
source the remaining part of the dilaton $\Pi_n$ and the spinor via
$a,b,c,d$.

Equation~\C{eqn:dil_susy_2}\ shows that the appearance of $(3,0)$
and non-primitive $(2,1)$ flux source the holomorphic part of the
dilaton, while~\C{eqn:dil_susy_1}\ shows the analogous statement for
the $(0,3)$ and $(1,2)$ components of $G_3$. The second pair of
equations,~\C{eqn:dil_susy_5}\ and~\C{eqn:dil_susy_6}, show that the
appearance of $(0,3)$ flux is related to the primitive and
non-primitive $(2,1)$-components of $G_3$, and vice-versa for the
$(3,0)$ component.

Thus, as noted in~\cite{DallAgata:2004dk}, we see that with $SU(2)$
structure, the three-form flux no longer need be $(2,1)$ and
primitive. Indeed, the non-primitive and $(3,0)$ parts result in a
non-holomorphic dilaton. We see both of these features explicitly in
our examples. Further, if we restrict to ISD fluxes, we are left
with a strict relation between the warp factor and the five-form
flux. Indeed, these equations determine a relation between specific
components of the three-form flux and the five-form flux which
sources the warp factor.

\subsubsection*{Gravitino:}

First consider the space-time component $M=\mu$ of~\C{eqn:sugra_2}.
We rewrite the gamma matrices in terms of the unwarped metric, and
use the fact the $\del_\mu \ve =0$ to find \be \delta\Psi_\mu =
-\frac{1}{4\kappa}\left[\gamma_\mu\otimes \gamma^n\del_n\log
e^{3\phi/2} \right]\ve + e^{3\phi/2}
\left[\gamma_\mu\otimes\gamma^n\del_n h\right]\ve +
\frac{e^{-3\phi/2}}{48}\left[\gamma_\mu\otimes\gamma^{npq}G_{npq}\right]\,\ve^*
=0.\non \ee We substitute the $\SU(2)$ spinor ansatz. For the
dilatino variation, the space-time component decouples giving two
independent equations proportional to $\zeta_+$ and $\zeta_-$: \bea
&&\left[-\frac{1}{4\kappa}\gamma^n\del_n \log e^{3\phi/2} +
e^{3\phi/4} \gamma^n \del_n h \right] (a \eta_- + b\chi_-) +
\frac{e^{-3\phi/2} }{48} \gamma^{npq}G_{npq} (c^* \eta_- + d^*
\chi_-) =
0,\non\\\non\\
&&\left[-\frac{1}{4\kappa}\gamma^n\del_n \log e^{3\phi/2} +
e^{3\phi/4} \gamma^n \del_n h \right] (c \eta_+ + d\chi_+) -
\frac{e^{-3\phi/2}}{48}  \gamma^{npq}G_{npq} (a^* \eta_+ + b^*
\chi_+) = 0.\non\\\non \eea We now contract with a complete basis of
spinors, as above, to give constraints on the warp factor and the
fluxes. As in the case of the dilaton, we may expand in terms of
$SU(2)$ invariants \bea
&&\del_n(\log e^{3\phi/2}) = \sigma w_n + \bar\sigma \bar w_n + \Sigma_n\non,\\
&&(\del_n h) = \theta w_n + \bar \theta \bar w_n + H_n.\non \eea
We
now get a series of equations
\bea ( \sigma - 4\kappa \theta
e^{3\phi/4}) a &=& \kappa e^{-3\phi/2} (i\wt
g_{(2,1)} c^* - 2g_{(1,2)} d^*),\label{eqn:grav_1}\\
(\bar \sigma + 4\kappa \bar \theta e^{3\phi/4})d &=& -\kappa
e^{-3\phi/2}\left(2 a^*  g_{(0,3)} - ib^* \wt
g_{(1,2)}\right),\label{eqn:grav_2} \eea and \bea &&(g-iJ)^{kn}
(\Sigma_n - 4\kappa e^{3\phi/4} H_n) a + K^{kn}(\Sigma_n - 4\kappa
e^{3\phi/4} H_n) b \non\\ && \qquad = -i \frac{\kappa}{2}
e^{-3\phi/2} \left[(g-iJ)^{kn} (V^1 - 2i V^2)_n c^* -
K^{kn}(V^1+2iV^2)_n
d^*\right],\label{eqn:grav_3}\\
&&(g+iJ)^{kn} (\Sigma_n + 4\kappa e^{3\phi/4} H_n) c + \bar
K^{kn}(\Sigma_n + 4\kappa e^{3\phi/4} H_n) d \non\\ && \qquad = i
\frac{\kappa}{2} e^{-3\phi/2} \left[(g+iJ)^{kn} (V^1 - 2i V^2)_n a^*
- \bar K^{kn}(V^1+2iV^2)_n b^*\right].\label{eqn:grav_4} \eea If we
restrict to ISD fluxes, we are left with a strict relation between
the warp factor and the five-form flux. This is to be expected from
the type B SUSY analysis.

In the specific examples we construct, this is not the case since
the flux is not ISD. Indeed, these equations determine a relation
between specific components of the three-form flux, and the
five-form flux which source the warp factor.

We now solve the internal component of the gravitino with $M=m$ in
\C{eqn:sugra_2}. This is the most involved calculation, and will
give a general equation determining the coefficients $a,b,c,d$ in
terms of the fluxes. Following the reasoning above, we find two
independent equations from $\zeta_+$ and $\zeta_-$: \bea && \nabla_m
(a\eta_-) + \nabla_m(b \chi_-) + \left(
\frac{1}{8}(\gamma_m^{\,\,\,n}-\delta_m^{\,\,\,n}) \del_n \log
e^{3\phi/2} -\frac{i}{2}Q_m\right)(a\eta_- + b\chi_-)\cr && = -
\frac{e^{3\phi/2}\kappa}{2}\gamma^n \gamma_m \del_n h (a \eta_- +
b\chi_-) +\frac{e^{-3\phi/4}\kappa}{96} \left(\gamma^{npq}_{\quad m}
+ 9\gamma^{[np}\delta^{q]}_{\,\,\,m}\right)G_{npq} (c^*\eta_- + d^*
\chi_-),\non\\
&&\nabla_m (c\eta_+) + \nabla_m (d\chi_+) + \left(
\frac{1}{8}(\gamma_m^{\,\,\,n}-\delta_m^{\,\,\,n})\del_n \log
e^{3\phi/2}
-\frac{i}{2}Q_m\right)(c\eta_+ + d\chi_+)\non\\
&& =  \frac{e^{3\phi/2}\kappa}{2}\gamma^n \gamma_m \del_n h (c
\eta_+ + d\chi_+) +\frac{e^{-3\phi/4}\kappa}{96}
\left(\gamma^{npq}_{\quad m} +
9\gamma^{[np}\delta^{q]}_{\,\,\,m}\right)G_{npq} (a^*\eta_+ + b^*
\chi_+),\non \eea which give two independent equations for
$\nabla_m\eta_-$ and $\nabla_m\chi_-$. Rewriting these equations
gives \bea \nabla_m \eta_- &=&  \frac{1}{\Delta}(d^* \del_m a -
b\del_m c^*) \eta_- + \frac{1}{\Delta}(d^* \del_m b - b\del_m
d^*)\chi_- +
\frac{1}{8}(\gamma_m^{\,\,\,n}-\delta_m^{\,\,\,n})\del_n \log
e^{3\phi/2} \eta_- -\cr && -\frac{i}{2} Q_m \eta_-
-\frac{e^{3\phi/2}\kappa}{2\Delta} \gamma^n \gamma_m \del_n h
\left[(ad^* + bc^*)\eta_- + 2bd^* \chi_-\right]\non\\&& +
\frac{e^{-3\phi/4}\kappa}{96\Delta} \left(\gamma^{npq}_{\quad m}+
9\gamma^{[np}\delta^{q]}_m\right)\left[(G_{npq}c^*d^* + \bar G_{npq}
ab)\eta_- + (G_{npq}d^{*\,2} + \bar G_{npq} b^2)\chi_-\right],\non
\eea and \bea \nabla_m \chi_- &=&  \frac{1}{\Delta}(c^* \del_m a -
a\del_m c^*) \eta_- + \frac{1}{\Delta}(c^* \del_m b - a\del_m
d^*)\chi_- -
\frac{1}{8}(\gamma_m^{\,\,\,n}-\delta_m^{\,\,\,n})\del_n \log
e^{3\phi/2} \chi_- +\cr && +\frac{i}{2} Q_m  \chi_-
+\frac{e^{3\phi/2}\kappa}{2\Delta} \gamma^n \gamma_m \del_n h
\left[2ac^*\eta_- + (bc^* + ad^*)\chi_-\right]\non\\&& -
\frac{e^{-3\phi/4}\kappa}{96\Delta} \left(\gamma^{npq}_{\quad m}+
9\gamma^{[np}\delta^{q]}_m\right)\left[(G_{npq}c^{*\,2} - \bar
G_{npq} a^2)\eta_- + (G_{npq}c^*d^* - \bar G_{npq}
ab)\chi_-\right],\non \eea where $\Delta = ad^* - bc^*$ is
nonsingular for $\SU(2)$ structure. Degenerate points are where the
structure becomes $\SU(3)$. We now use $d(\eta_+^T \eta_-)=0$,
$d(\chi_+^T \chi_-)=0$ and $d(\chi_+^T \eta_-)=0$ to give a series
of equations determining $a,b,c,d$.
\begin{enumerate}
\item Using $d(\eta_+^T \eta_-) = 0$ gives
\bea &&\frac{d^* \del_m a - b\del_m c^*}{\Delta\kappa} +
\frac{d\del_m a^* - b^* \del_m c}{\Delta^*\kappa} \non\\ &&=
-\frac{1}{8} \del_m \log e^{3\phi/2} -\frac{e^{3\phi/2}}{2}
(\delta_m^{\,\,\,n}-iJ_m^{\,\,\,n})\del_n h \frac{ad^* +
bc^*}{\Delta} - e^{3\phi/2} K_m^{\,\,\,n} H_n
\left(\frac{bd^*}{\Delta}\right)  \non\\
&& + \frac{e^{-3\phi/4}}{4\Delta}w_m\left[  4 g_{12} (d^*)^2 - ic^*
d^* \wt g_{21} -i ab \,\wt g_{12}^* + 4b^2 g^*_{21}\right] 
\non\\&&+ \frac{e^{-3\phi/4}}{4\Delta^*}w_m\left[  4 g^*_{03} (d)^2
+2 ic d \wt g^*_{12} +2i a^*b^* \, \wt g_{21} + 4(b^*)^2
g_{30}\right]\non\\&&
+\frac{e^{-3\phi/4}}{16\Delta}\left[(J_m^{\,\,\,n}-3i\delta_m^{\,\,\,n})
(V^1+2iV^2)_n c^* d^* + iK_m^{\,\,\,n} (3V^1 -
2iV^2)_n(d^*)^2\right]\non\\&&
+\frac{e^{-3\phi/4}}{16\Delta}\left[
(J_m^{\,\,\,n}-3i\delta_m^{\,\,\,n})(V^1+2iV^2)_n ab +
iK_m^{\,\,\,n}(3 V^1-2iV^2)_nb^2\right] + \cc
\label{eqn:susy_gravitino_1} \eea

\item  Now we use $d(\chi_+^T\chi_-)=0$. This gives
\bea
&&\frac{c^* \del_m b - a\del_m d^*}{\Delta \kappa} + \frac{c \del_m b^* - a^*\del_m d}{\Delta^* \kappa}\non\\
&&= -\frac{1}{8} \del_m \log e^{3\phi/2}
+\frac{e^{3\phi/2}}{2\Delta} \left[ (bc^* +
ad^*)\left[(\delta_m^{\,\,\,n}+iJ_m^{\,\,\,n}) \del_n h -
2(w_m \bar\theta - \bar w_m \theta)\right]\right]  \non\\
&& -\frac{ac^* e^{3\phi/2}}{\Delta} \bar K_m^{\,\,\,n} H_n
+\frac{e^{-3\phi/4}}{16\Delta} \left\{-(c^*)^2\bar K_m^{\,\,\,n}
(2V^2+3iV^1)_n  \right\}\cr && +\frac{e^{-3\phi/4}}{16\Delta}
\left\{ a^2 K_m^{\,\,\,n} (2V^2+3iV^1)_n -
(J_m^{\,\,\,n}+3i\delta_m^{\,\,\,n}) (V^1+2iV^2)_n(c^*d^*)\right\} 
\cr &&+\frac{e^{-3\phi/4}}{16\Delta}\left\{
(ab)(J_m^{\,\,\,n}+3i\delta_m^{\,\,\,n})(V^1-2iV^2)_n + 12
 V^2_n(ab-c^*d^*)\right\} \non\\
 &&+\frac{e^{-3\phi/4}}{4\Delta} w\left[2g_{30}(c^*)^2 -
 2g^*_{03} a^2 - 2i \wt g_{21}c^*d^*
 +2i \wt g_{12}^*ab\right]  \non\\ && + \frac{e^{-3\phi/4}}{4\Delta^*} w
\left[2g_{21}^* c^2 - 2g_{12}(a^*)^2 + i\wt g_{12}^* (cd) - i \wt
g_{21}(ab)^*\right]+\cc\label{eqn:susy_gravitino_2} \eea

\item Finally we obtain a complex equation using
$d(\chi_+^T\eta_-)=0$. Using the above contractions, we get \bea &&
\frac{b \del_m d^* - d^* \del_m b}{\Delta} + \frac{a^* \del_m c -
c\del_m a^*}{\Delta^*} = -\frac{1}{8} (K + \bar
K)_m^{\,\,\,n}\Sigma_n\non\\&& + \frac{e^{3\phi/2}\kappa}{2}
\left[K_m^{\,\,\,n} H_n \frac{ad^* + bc^*}{\Delta} -\bar
K_m^{\,\,\,n} H_n \frac{b^*c + a^*d}{\Delta^*}\right] \non\\ && -
e^{3\phi/2} \kappa \left[ \left(\frac{bd^*}{\Delta}+ \frac{a^*
c}{\Delta^*}\right) (\delta_m^{\,\,\,n}+iJ_m^{\,\,\,n})\del_n h +
\frac{2bd^*}{\Delta}
w_{[m} \bar w_{n]} \del^n h\right]\non\\
&& +\frac{e^{-3\phi/4}\kappa}{16} \left[i \bar K_m^{\,\,\,n}(3V^1 -
2iV^2)_n \left(\frac{c^*d^* + ab}{\Delta}\right) + i \bar
K_m^{\,\,\,n}(3V^1 + 2iV^2)_n \left(\frac{cd -
a^*b^*}{\Delta^*}\right)\right.\non\\&& \left.
+(J_m^{\,\,\,n}+3i\delta_m^{\,\,\,n})(V^1-2iV^2)_n\left(\frac{(d^*)^2}{\Delta}
+ \frac{(a^*)^2}{\Delta^*}\right) \right.\cr
&&\left.+(J_m^{\,\,\,n}+3i\delta_m^{\,\,\,n})(V^1+2iV^2)_n\left(\frac{b^2}{\Delta}
- \frac{(c^*)^2}{\Delta^*}\right)
\right].\label{eqn:susy_gravitino_3} \eea As opposed to the previous
two real equations,~\C{eqn:susy_gravitino_1}\
and~\C{eqn:susy_gravitino_2}, this is a complex equation. We
therefore find four independent real equations determining the
coefficients $a,b,c,d$.
\end{enumerate}
These results can be used to check that the type IIB supersymmetry
variations vanish for our explicit examples. As an additional check,
we used Mathematica to check that the equations of motion are
satisfied.

\newpage
\ifx\undefined\bysame
\newcommand{\bysame}{\leavevmode\hbox to3em{\hrulefill}\,}
\fi

\end{document}